# Probing the roles of orientation and multi-scale gas distributions in shaping the obscuration of Active Galactic Nuclei through cosmic time

Alba V. Alonso-Tetilla[1]★, Francesco Shankar[1]†, Fabio Fontanot[2,3], Nicola Menci[4], Milena Valentini[5,6,2], Johannes Buchner[7], Brivael Laloux[8,9], Andrea Lapi[10], Annagrazia Puglisi[1,11]‡, David M. Alexander[9], Viola Allevato[12], Carolina Andonie[9], Silvia Bonoli[19,20], Michaela Hirschmann[11,2], Iván E. López[13,14], Sandra I. Raimundo[15,16,1], Cristina Ramos Almeida[17,18]

[1] *School of Physics and Astronomy, University of Southampton, Highfield, SO17 1BJ, Southampton, UK*
[2] *INAF - Astronomical Observatory of Trieste, via G.B. Tiepolo 11, I-34143 Trieste, Italy*
[3] *IFPU – Institute for Fundamental Physics of the Universe, via Beirut 2, I-34151 Trieste, Italy*
[4] *INAF - Osservatorio Astronomico di Roma, via Frascati 33, I-00078 Monteporzio, Italy*
[5] *Astronomy Unit, Department of Physics, University of Trieste, via Tiepolo 11, I-34131 Trieste, Italy*
[6] *Universitats-Sternwarte Munchen, Fakultat fur Physik, LMU Munich, Scheinerstr. 1, 81679 Munchen, Germany*
[7] *Max Planck Institute for Extraterrestrial Physics, Giessenbachstrasse, 85741 Garching, Germany*
[8] *Institute for Astronomy & Astrophysics, National Observatory of Athens, V. Paulou & I. Metaxa, 11532, Greece*
[9] *Centre for Extragalactic Astronomy, Department of Physics, Durham University, UK*
[10] *SISSA, Via Bonomea 265, 34136 Trieste, Italy*
[11] *Institute for Physics, Laboratory for galaxy Evolution and Spectral Modelling, EPFL, Observatoire de Sauverny, Chemin Pegasi 51, 1290 Versoix, Switzerland*
[12] *INAF-Osservatorio Astronomico di Capodimonte, Via Moiariello 16, 80131 Napoli, Italy*
[13] *Departimento di Fisica e Astronomia "Augusto Righi", Universita di Bologna, Via Gobetti 93/2, 40129 Bologna, Italy*
[14] *INAF - Osservatorio di Astrofisica e Scienza dello Spazio di Bologna, Via Gobetti 93/3, 40129 Bologna, Italy*
[15] *Department of Physics and Astronomy, University of California, Los Angeles, 90095, California, USA*
[16] *DARK, Niels Bohr Institute, University of Copenhagen, Jagtvej 155, Copenhagen N, 2200, Denmark*
[17] *Instituto de Astrofísica de Canarias, Calle Vía Láctea, s/n, E-38205, La Laguna, Tenerife, Spain*
[18] *Departamento de Astrofísica, Universidad de La Laguna, E-38206, La Laguna, Tenerife, Spain*
[19] *Donostia International Physics Center, Paseo Manuel de Lardizabal 4, E-20118 Donostia-San Sebastián, Spain*
[20] *Ikerbasque, Basque Foundation for Science, E-48013 Bilbao, Spain*

**ABSTRACT**

The origin of obscuration in Active Galactic Nuclei (AGN) is still an open debate. In particular, it is unclear what drives the relative contributions to the line-of-sight column densities from galaxy-scale and torus-linked obscuration. The latter source is expected to play a significant role in Unification Models, while the former is thought to be relevant in both Unification and Evolutionary Models. In this work, we make use of a combination of cosmological semi-analytic models and semi-empirical prescriptions for the properties of galaxies and AGN, to study AGN obscuration. We consider a detailed object-by-object modelling of AGN evolution, including different AGN light curves (LCs), gas density profiles, and also AGN feedback-induced gas cavities. Irrespective of our assumptions on specific AGN LC or galaxy gas fractions, we find that, on the strict assumption of an exponential profile for the gas component, galaxy-scale obscuration alone can hardly reproduce the fraction of $\log(N_{\rm H}/{\rm cm}^{-2}) \geq 24$ sources at least at $z \lesssim 3$. This requires an additional torus component with a thickness that decreases with luminosity to match the data. The torus should be present in all evolutionary stages of a visible AGN to be effective, although galaxy-scale gas obscuration may be sufficient to reproduce the obscured fraction with $22 < \log(N_{\rm H}/{\rm cm}^{-2}) < 24$ (Compton-thin, CTN) if we assume extremely compact gas disc components. The claimed drop of CTN fractions with increasing luminosity does not appear to be a consequence of AGN feedback, but rather of gas reservoirs becoming more compact with decreasing stellar mass.

**Key words:** Galaxies: active - galaxies: formation - galaxies: evolution - galaxies: fundamental parameters - quasars: supermassive black holes - black hole physics – galaxies: nuclei – galaxies: structure.

# 1 INTRODUCTION

The study of Active Galactic Nuclei (AGN[1]) obscuration is one of the most crucial but still largely debated issues to fully characterize

★ E-mail: A.V.Alonso-Tetilla@soton.ac.uk
† E-mail: F.Shankar@soton.ac.uk
‡ Anniversary Fellow

© 2023 The Authors

---

[1] In this paper, we define as AGN all massive galaxies hosting a supermassive black hole (SMBH) at their centre powered by gas accretion which has had

AGN demography and shed light on the cosmological evolution of supermassive black holes (SMBHs). An AGN is defined as obscured when the emission from the accretion disc (UV and optical wavelengths) is blocked by intervening absorbing material along the line of sight (e.g., Seyfert 1943; Antonucci 1993; Urry & Padovani 1995; Netzer 2015).

Obscuration in AGN can originate due to the presence of gas and dust in the interstellar medium of the host galaxy (Lapi et al. 2005; Buchner et al. 2017; Gilli et al. 2022, and references therein), and/or due to an intervening inner dusty *torus* at a distance of a few parsecs from the central SMBH (e.g., Packham et al. 2005; Radomski et al. 2008; Burtscher et al. 2013; Imanishi et al. 2016; García-Burillo et al. 2016; Gallimore et al. 2016). The torus is a dynamical and clumpy structure that might be part of the dusty wind originating from the accretion disc (e.g., Ramos Almeida et al. 2009, 2011; Wada 2012; Markowitz et al. 2014; López-Gonzaga et al. 2016; Hönig & Kishimoto 2017). The presence of gas and dust in a galaxy varies along its lifetime, being more abundant during the early formation phases of the galaxy, whilst decreasing at later times (Granato et al. 2004; Santini et al. 2014; Lapi et al. 2006, 2014). In original studies, a torus component around a SMBH may be a long-lived structure (Urry & Padovani 1995). However, some studies suggest that this dynamical structure most likely appears and disappears with the periods of nuclear activity (e.g., Ramos Almeida & Ricci 2017; García-Burillo et al. 2019). Unveiling the nature of obscuration in AGN can thus not only provide a more complete census of AGN/SMBH through cosmic time (e.g., Shankar et al. 2013), but also set valuable constraints on the overall co-evolution of SMBHs with their host galaxies. This evolution could be a result of obscured AGN preferentially living in the early growth phases of the accretion evolution of the central SMBH and its host galaxy (Sanders et al. 1989; Granato et al. 2004; Lapi et al. 2006; Hopkins et al. 2007), or instead be the result of a mere orientation effect (Urry & Padovani 1995; Polletta & Courvoisier 1999). Given the considerations above, two distinct, though not necessarily mutually exclusive, scenarios have been put forward in the literature to explain obscuration in AGN. The first one, known as the Unification or Orientation model (Antonucci 1993; Urry & Padovani 1995; Netzer 2015), proposes that most of the obscuration in AGN is a consequence of the orientation of the observer with respect to the SMBH, or the orientation between the AGN and its torus. Observing a galaxy and/or its central SMBH edge-on, would clearly amplify the line-of-sight obscuration due to gas and dust in the host galaxy and the torus. The second model, known as Evolutionary model, supports the idea that the level of obscuration, AGN luminosity, and SMBH mass, all depend on the evolutionary stage of the host galaxy, largely irrespective of the observer line-of-sight. In traditional evolutionary models, the SMBH grows in a two-mode fashion, (super)Eddington-limited initially, until it reaches a peak luminosity (possibly regulated by AGN feedback effects), followed by a usually longer sub-Eddington phase. In the *pre-peak* phase, the AGN is considered to be obscured, whilst during the *post-peak* phase the SMBH mass has grown sufficiently enough to expel and/or ionize cold gas in the host galaxy via the feedback action of AGN winds and/or jets (Granato et al. 2004; Brandt & Hasinger 2005; Granato et al. 2006; Lapi et al. 2006; Hopkins et al. 2007). Some theoretical models also showed that it is possible to reproduce the X-ray AGN luminosity functions of both obscured and unobscured AGN by considering the former sources shining *pre-peak* and the latter *post-peak* in the AGN light curve, i.e., occurring preferentially before or after the AGN feedback blowout phase (e.g., Granato et al. 2006; Lapi et al. 2006; Hopkins et al. 2008). However, despite the numerous attempts, a focused modelling on AGN obscuration from pure orientation models in a full cosmological context is still missing (but see the work by Lapi et al. 2006; Menci et al. 2008; Gilli et al. 2022, for some initial attempts in this direction). In this work, we specifically focus on orientation-driven AGN obscuration, although we will also briefly discuss models which include a torus component that may exist for only part of the lifetime of an AGN.

From the observational perspective, it is clear that both orientation and evolutionary components are in action in AGN, possibly with varying importance depending on time, mass, and environment of the host galaxy. The absence of broad emission lines in the so-called *Type 2* AGN (for definitions and *Type* differences see Antonucci 1993), for example, is usually interpreted as an orientation effect from evidence in polarized spectra. The broad emission lines originating from the central region/clouds, are suppressed when line-of-sight is sufficiently edge-on and covered by the torus, although some sources may have intrinsically weaker broad emission lines (e.g., naked AGN, Hawkins 2004).

On statistical grounds, there is often good correspondence in the AGN luminosity functions of X-ray selected sources with $\log(N_H/\text{cm}^{-2}) < 22$, and *Type 1* AGN from optical/UV surveys (e.g., Ueda et al. 2003; Ricci et al. 2017a), or in the Eddington rate distribution (e.g., Ricci et al. 2017b). On the other hand, the host galaxy evolution must also play a role in AGN obscuration. For example, some studies have found that *Type 2* AGN are less massive than *Type 1*, disagreeing with the Orientation model stating that AGN *Type* 1 and 2 are the same objects observed along different viewing angle (Ricci et al. 2022). Alexander et al. (2005) showed that many of the SCUBA-detected sources host luminous X-ray AGN, and more recently obscured AGN seem to be ubiquitous in starbursts and also more regular star-forming galaxies (e.g., Mountrichas & Shankar 2023, and references therein).

The relation between star formation, obscuration and fuelling mechanisms of AGN and the connection between them has been a topic of significance in the community. Some studies find strong correlation between the column density and the presence of a stellar bar in Type 2 galaxies (Maiolino et al. 1999), while others find a power-law relation between column density and stellar mass in long-duration gamma ray bursts (Buchner et al. 2017). Buchner & Bauer (2017) found that the galaxy-scale gas is responsible for a luminosity-independent fraction of $N_H \sim [10^{22}\text{--}10^{24}]$ cm$^{-2}$ AGN obscuration but does not produce $N_H > 10^{24}$ cm$^{-2}$, suggesting that observations like Ueda et al. (2014) or Ananna et al. (2019), where the fractions of $N_H \sim [10^{22} - 10^{24}]$ cm$^{-2}$ present a luminosity dependency, is due to the luminosity dependency from the torus component rather than the galaxy-scale obscuration. Whitaker et al. (2017) observed a strong dependence of the dust attenuation obscuration with stellar mass with a small redshift evolution ($z = 0 - 2.5$), so an unobscured-to-obscured phase could mean a transition is happening for low stellar masses (see also Kashino et al. 2013; Pannella et al. 2015; Reddy et al. 2015; Shivaei et al. 2015, and references therein).

A variety of methods are adopted to identify and characterize obscured AGN (for an extensive overview of the different methods depending on wavelength see Hickox & Alexander 2018). X-ray observations (e.g., Giacconi 2009) are one of the best methods for selecting obscured AGN since they are directly associated with the accretion disc, and its hot corona. X-rays have more penetrating power through thick mediums, at least until the Compton-thin/thick limit of $N_H \sim 10^{24}$ cm$^{-2}$. A variety of observational studies have

---

a bolometric luminosity above $L_{\text{bol}} > 10^{42}$ erg/s at some point during their active lives.

attempted to describe the demography and evolution of AGN as a function of their column densities (e.g., Ueda et al. 2014; Aird et al. 2015; Buchner et al. 2015; Ananna et al. 2019; Laloux et al. 2023). It has been several times recognized that the Cosmic X-ray Background (CXB) of AGN can be reproduced by a collection of AGN with varying column densities ranging from $N_H \sim 10^{20} - 10^{26}$ cm$^{-2}$ (e.g., Gilli et al. 2007; Shen 2009; Shankar et al. 2009; Ueda et al. 2014; Aird et al. 2015; Ananna et al. 2019; Gilli et al. 2022, and references therein).

In this paper, we use the comprehensive semi-analytic model (SAM) for GAlaxy Evolution and Assembly (GAEA, Fontanot et al. 2020, F20 hereafter) as a self-consistent baseline for a realistic simulated population of galaxies and their central SMBHs, consistent with the present constraints on the galaxy stellar mass function and AGN luminosity function. Starting from GAEA predictions, we then assign to each model galaxy a line-of-sight Hydrogen column density $N_H$, based on its gas mass, as well as a torus component based on its SMBH mass and AGN luminosity. However, we also check the robustness of our results by varying various key prescriptions of the GAEA model in a semi-empirical fashion, by adopting, for example, different AGN light curves, gas fractions, or gas disc sizes.

This paper is the first of a pair dedicated to the study of the origin of obscuration in AGN. This paper is dedicated to the Orientation model, while its companion paper (Alonso-Tetilla et al. in preparation) will focus on Evolutionary models. The plan of the paper is a follows. We present a detailed description of the adopted methodology in Section 2, where we describe the basic information produced by GAEA which we use as baseline for our calculations, the computation of column density from the large-scale gas distribution, the inclusion of an AGN-driven Blast Wave, and the modelling of a dusty torus-like central component. In Section 3 we present our main results in terms of the key model parameters driving AGN obscuration in our orientation-based model. We discuss our results in Section 4, where we highlight the impact of varying any of our underlying assumptions or parameters, and we then list our main conclusions in Section 5.

## 2 METHODOLOGY

Our methodology to study the statistical distribution of obscured and unobscured AGN in orientation models relies on the following steps:

(i) We start from a realistic mock of galaxies (GAEA galaxy catalogues) at a given redshift consistent with available data on the stellar mass function and AGN/quasar (QSO) luminosity functions.

(ii) We then assign to each galaxy a HI line-of-sight column density based on its gas content and geometry, and examine the effect of an AGN-driven Blast Wave in modulating the $N_H$.

(iii) To each galaxy, we also assign a torus-like component based on its SMBH mass, and AGN luminosity.

(iv) We then repeat the steps above at different epochs to study the predicted evolution of AGN obscuration as a function of redshift.

As previously mentioned, we use as a reference the galaxies and SMBHs extracted from the GAEA SAM, which also yields cold gas fractions, disc sizes, and SMBH accretion rates (light curves). The advantage of using this SAM is that a state-of-the-art cosmological model provides inner self-consistency among the different variables and models used, for example retaining the AGN feedback-induced relation between gas fractions and AGN luminosity. Nevertheless, in a data-driven approach, we also explore the impact on our results by varying, in turn, galaxy gas fractions, AGN light curves, and galaxy radii as guided by observational results. We show that our main results are broadly invariant under these changes except for some notable examples which we discuss in detail in the next Sections.

The GAEA model is described in Section 2.1, we provide full details on how we compute galaxy-scale obscuration in Section 2.2, while in Section 2.3 we discuss how we assign a torus-like component to each active galaxy.

### 2.1 GAEA

In this paper we present a study of orientation-driven (Hydrogen) obscuration in AGN in a cosmological context taking advantage of the predictions of the semi-analytic model GAEA (F20), which follows the evolution of galaxies and their central SMBHs from early times down to the present epoch. GAEA follows state-of-the-art recipes to describe the evolution of stars and gas in galaxies, as well as providing a detailed modelling of the growth of the central SMBHs. We hereby provide a brief overview of GAEA's modelling of SMBHs, while full details can be found in F20. In particular, in this paper we focus on the so-called HQ11-GAEA realization, which includes Hopkins & Quataert (2011) and Hopkins et al. (2006) prescriptions to estimate:

(i) The fraction of the cold gas available in the host galaxy which loses enough angular momentum to reach the central regions and accumulate into a low angular momentum gas *reservoir*.

(ii) The accretion onto the SMBH from material accumulated into the reservoir or accretion disc, in particular the accretion rate follows a fixed AGN light curve based on the results of the numerical hydro-simulations.

The original model has been calibrated on Dark Matter Merger trees drawn from the Millennium Simulation (Springel et al. 2005), which typically allows for a good description of galaxy properties down to a stellar mass scale of the order of $10^9$ M$_\odot$.

SMBH seeding in GAEA is performed following Volonteri et al. (2011) and corresponds to seed masses of $\sim 10^4$ M$_\odot$ (which is the resolution of the Millennium Simulation). The subsequent growth of these seeds is then followed via gas accretion and mergers with other SMBHs. The accretion of gas onto the SMBH in GAEA is triggered by both galaxy mergers and disc instabilities, which contribute to the creation of a central gas reservoir of low angular momentum, which in turn gradually feeds the central SMBH. The accretion onto the central SMBH is then redistributed in time following an AGN light curve, namely composed of an initial (super-)Eddington accretion phase, which lasts until the SMBH reaches the self-regulation limit, followed by a power-law decline, as also suggested by theoretical arguments and hydrodynamic simulations (e.g., Granato et al. 2004; Lapi et al. 2006; Hopkins et al. 2007; Shen 2009). GAEA, as well as radio-mode feedback, also includes QSO-mode feedback in the form of winds. AGN winds heat the cold gas eventually expelling it in the hot gas. Specifically, the model realization considered in this work, HQ11-GAEA, uses the outflow rate predictions as a function of cold gas mass, bolometric luminosity and black hole mass from Menci et al. (2019).

The HQ11-GAEA model is calibrated to reproduce the evolution of the AGN luminosity function without applying any obscuration correction to model predictions, while still reproducing all galaxy properties discussed in previous papers (e.g., Hirschmann et al. 2016), like mass-metallicity relations, quenched fractions and cold gas fractions. It also reproduces the observed distribution of Eddington ratios at various redshifts (F20). A deeper analysis on the chemical enrichment can be found in De Lucia et al. (2014).

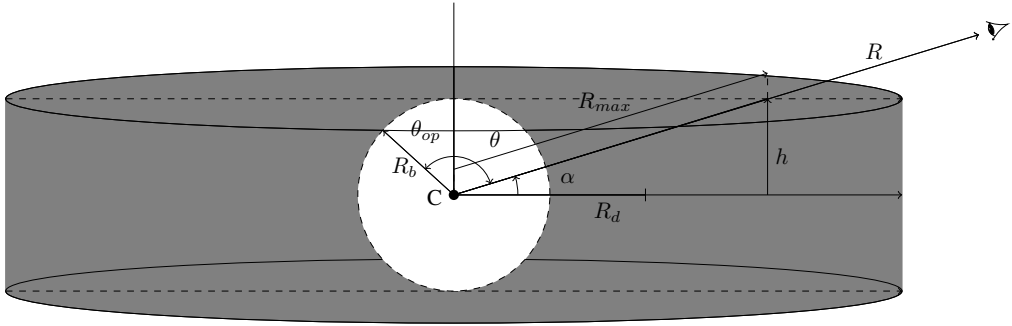

**Figure 1.** Diagram depicting a typical galaxy in our mock sample characterised by an exponential gas density profile and a disc geometry. In this figure, $R$ traces the line-of-sight, $R_d$ is the gas scale length, $h$ is the thickness of the disc, $\theta$ is the angle between the vertical and $R$, $\alpha$ is the random angle for the random line-of-sight, $R_{max}$ marks the end point of the galaxy edges along $R$, $R_b$ is the radius of the Blast Wave, and $\theta_{op}$ is the opening angle of the Blast Wave (central white sphere in the Figure).

### 2.2 Column density distribution: contribution from the galaxy

Throughout this work we consistently assume that the gas density in star-forming discs follows an exponential density profile,

$$\rho(R, \theta) = \rho_0 \exp(-R/R_d), \quad (1)$$

where $\rho_0$ is the central gas volume density, $R$ is the line-of-sight radius from the centre of the disc to the furthermost part of the galaxy, $R_d$ is the gas disc scale length, and $\theta$ is the angle between the rotational vertical axis of the galaxy and the line of sight. The geometry is visualized in Figure 1.

Although possibly not all galaxies are characterized by exponential density profiles for their gas component (e.g., van der Kruit 1979; Pohlen & Trujillo 2006; Bigiel & Blitz 2012; Wang et al. 2014), Eq. (1) still represents a good approximation to the gas mass distribution of many galaxies at different epochs and stellar masses, and becoming even a better approximation at higher redshifts (e.g., Patterson 1940; Freeman 1970; Hodge et al. 2019; Hunter et al. 2021; Ferreira et al. 2022, and references therein). In some highly star-forming high-redshift galaxies, the gas disc profile may deviate from an exponential one, attaining a more compact and spherical geometry. We will anyhow continue adopting an exponential gas density profile, although we will also briefly discuss the impact of switching to, e.g., a Sersic profile in Section Discussion, as well as in our companion paper. Also, Eq 1 is consistent with the fact that GAEA, following common recipes in SAMs, assumes that the cold gas density settles in an exponential profile once in rotational equilibrium. This gas profile is consistent with surface brightness profiles observed in some galaxy samples (e.g., ASPECS, Aravena et al. 2020, and CANDELS, Grogin et al. 2011; Koekemoer et al. 2011, or Hodge et al. 2019), as predicted by hydrodynamic simulations (Aumer et al. 2013), and used by previous semi-analytic models (Fu et al. 2009, 2010), but with a gas disc scale length that could be different from the stellar component, as discussed below.

The normalization $\rho_0$ in Eq. (1) is chosen in a way that the integral of the gas density over the full radial ($R$), vertical ($\theta$) and azimuthal ($\phi$) extent of the galaxy equals the gas mass of the host galaxy ($M_{gas}$), i.e.,

$$\int_0^{2\pi} \int_0^{\pi} \int_0^{\infty} \rho(R,\theta) R^2 \sin\theta \, dR \, d\theta \, d\phi = M_{gas}$$

$$\Rightarrow \rho_0 = \frac{M_{gas}}{4\pi R_d^2 I}, \quad (2)$$

where

$$I = \int_0^{\infty} e^{-x} x^2 dx = 2, \quad (3)$$

and $x = R/R_d$. For simplicity of visualization and computation, especially in models with a Blast Wave, we will always use cylindrical coordinates in all our calculations.

Throughout the paper, we assume a disc thickness $h = R_d/8$ (e.g., Nath Patra 2020). As a sanity check, we also consider other possible definitions (e.g., Ojha 2001, $h = R_d/15$) showing that the disc thickness plays a relatively minor role in AGN obscuration with respect to other input variables. This test is discussed in Appendix A.

The line-of-sight HI column density is then calculated as

$$N_H = \int_0^{R_{max}} \rho(R,\theta) dR = \int_0^{\frac{h}{R_d \sin\alpha}} \rho_0 e^{-x} dx, \quad (4)$$

where $\alpha$ is the angle between the plane of the galaxy and the line-of-sight, and $R_{max}$ is maximum radius of the galaxy for a given line-of-sight (see Figure 1).

We are making two assumptions in this methodology:

- What X-ray observations measure is driven by hydrogen ($N_H$ from X-ray spectra assume solar abundance of H/He/O/Fe).
- That $M_{gas}$ consists of 100% hydrogen.

As mentioned before, GAEA includes the effects of AGN feedback on reducing the gas mass in the low angular momentum reservoir around the SMBH, as well as in the surrounding galaxy. However, GAEA does not include the dynamical effect of AGN feedback on the gas distribution in the host galaxy, which becomes relevant when calculating the line-of-sight $N_H$ column density. To include this effect, we follow a model based on the AGN-driven outflows proposed by Menci et al. (2019) (see also Lapi et al. 2005; Menci et al. 2008), which analytically follows the two-dimensional expansion of AGN-driven outflows as a function of the global properties of the host galaxy and of the luminosity of the central AGN. These AGN driven outflows are effectively *winds* (for a review see King & Pounds 2015). Different theoretical works (e.g., Silk & Rees 1998; King 2003; Granato et al. 2004; Lapi et al. 2005; Silk & Nusser 2010; King et al. 2011; Faucher-Giguère & Quataert 2012) have tried to capture the main features of the outflows using models based on shocks expanding into the interstellar medium (ISM). These models use power-law density profiles or exponential discs (Hartwig et al. 2018), and spherical approximation, consistent with AGN-driven outflows with luminosity, outflow rate, and shock velocity dependence. The Menci et al. (2019) treatment effectively follows the expansion velocity of the shock and the mass outflow rate out to large radii where the stellar/gas discs are the dominant components, creating a Blast Wave (BW). This BW creates a cavity in the gas, pushing the gas to the outskirts of the galaxy and creating a thin layer around the BW with heavily compressed gas. Menci et al. (2019) provide tabulated numerical solutions for the fraction of the cold gas ejected ($f_{qw}$) as a function of the properties of the AGN and host galaxy. These scaling have been included in the F06-GAEA SAM (F20), showing that AGN-driven winds help in reproducing the (low) levels of SFR in massive galaxies, removing some cold gas still in place in these galaxies since $z \sim 2$.

We compute the opening angle of the AGN BW for each galaxy which depends on the AGN bolometric luminosity, the cold gas mass $M_{gas}$, and the virial velocity of the parent Dark Matter Halo $V_{vir}$ following the tabulated values by Menci et al. (2019). In this work, the virial velocity is used to predict the amount of reduction in gas mass in the centre of galaxies, and its impact on the line-of-sight column density. This model assumes that the opening angle corresponds to the maximum aperture of the BW, which occurs at the peak luminosity of the AGN. Therefore, our $N_H$ corresponds to the one at the maximum value of the bolometric luminosity, although we also explore model variants where we relax this assumption.

Assuming gas mass conservation during the expansion of the BW (Lapi et al. 2005; Menci et al. 2008), the part of the total gas mass that is pushed away by the bubble creating a central cavity will all be compressed in a thin layer around the bubble. When $R_b > h$, the BW pushes the gas outside the disc galaxy and part of the gas is removed. In that scenario, a line-of-sight with $\alpha \sim 0$ will see the same $N_H$ as without the BW. However, when $R_b > R_{max}$, the BW removes the gas from the line-of-sight reducing $N_H$. We can thus compute the total line-of-sight column density $N_H$ as the sum of two components, the contribution from the shell and from the outer, still unperturbed gas disc

$$N_H = \left(\frac{\theta_{op}}{90}\right) N_H^{shell} + N_H^{out} \qquad (5)$$

The column density of the outside, unperturbed disc is

$$N_H^{out} = \rho_0 \frac{Q_{out}}{I} \qquad (6)$$

where $I$ is given in Eq 3 and $Q_{out}$ is defined as

$$Q_{out} = \int_{R_b/R_d}^{R_{max}/R_d} x^2 e^{-x} dx, \quad R_b < R_{max} \qquad (7)$$

with $R_b$ the radius of the bubble calculated from $\theta_{op}$. The column density contribution from the shell is instead given by

$$N_H^{shell} = \frac{M_{gas}^{shell}}{4\pi R_b^2} \qquad (8)$$

where the cold gas mass of the cavity is calculated by

$$M_{gas}^{shell} = M_{gas} \frac{Q_{shell}}{I} \quad \text{and} \quad Q_{shell} = \int_0^{R_b/R_d} e^{-x} x^2 dx. \qquad (9)$$

In our reference model, we assume that the column density does not evolve during the lifetime of the AGN. However, AGN feedback models predict some evolution in the amount of gas content in the host galaxy already during the relatively brief lifetime of the AGN (see, e.g., Granato et al. 2004; Lapi et al. 2006, 2014; Santini et al. 2014). Indeed, even in GAEA the gas mass is reduced by the AGN feedback. However, Eq 1 considers a single snapshot of $M_{gas}$ when the AGN is at the beginning of its light curve. The gas mass reaches a maximum value at the start of the SMBH active phase, and then rapidly decreases around and after the peak of the light curve (see also Cavaliere et al. 2002; Lapi et al. 2005). This evolution is explored in Appendix B. In what follows, for convenience all our main results are plotted against the peak luminosity (luminosity at the peak accretion rate within the light curve), although we will show in Appendix B that this assumption plays a minor role on our results.

The bolometric luminosities are directly calculated from the gas accretion rates onto the central SMBHs. More specifically, the GAEA model includes both a QSO- and a radio-mode AGN feedback, each one characterized by its own independent radiative efficiency which sets the fraction of rest mass energy of the accretion flow onto the SMBH that is converted into radiative or kinetic luminosity, respectively. The radio-mode feedback is generally less efficient, with a kinetic efficiency of just 2%, against the 15% assumed for the radiative-mode feedback. In Fontanot et al. (2020), both the contributions of the QSO- and Radio-mode accretion have been taken into account to estimate the AGN/QSO luminosity function (LF). In general, for consistency, we follow the same approach. It is worth stressing that Radio-mode accretion becomes relevant only for massive galaxies residing in massive haloes at low-redshifts, as those are the environments where an efficient quenching of the cooling flows and late SFR is required. Radio-mode accretion, by construction, is treated as an (almost) continuous accretion process of hot gas from the halo (which gives rise to tensions with the observed distribution of radio galaxies - see e.g., Fontanot et al. 2011). At low redshift, this implies that the radio-mode is dominant in galaxies devoid of their cold gas content. On the other end, in modelling the QSO-mode, GAEA is explicitly dealing with the flow of the cold gas from the host galaxy disc to the reservoir, and with the effects of feedback on the evolution of the total cold gas content. These considerations imply that our geometrical modelling of obscuration correlates better with the QSO-mode prescription, while the Radio-mode channel is typically underestimating the obscuration by construction. We will thus also present model predictions on the AGN obscured fractions removing the sources dominated by Radio-mode accretion and show that these are very similar to the full model outputs at z> 2, but diverge somewhat at low z and low L, as further detailed below. The 2-10 keV intrinsic X-ray luminosities are calculated from bolometric luminosities via the bolometric correction by Duras et al. (2020). Similar results would be retrieved adopting, for example, the Marconi et al. (2004) bolometric correction.

### 2.3 Column density distribution: contribution from the *torus*

It is now clear from direct and indirect (via, e.g., Spectral Energy Distribution (SED) fitting) observations that a torus-like component (Combes et al. 2019; García-Burillo et al. 2019, 2021) is an essential ingredient required to fully model the observational properties of AGN (see Netzer 2015; Ramos Almeida & Ricci 2017; Hickox & Alexander 2018, for reviews), especially in their $\log_{10}(N_H/cm^{-2}) > 24$ phase (Risaliti et al. 1999; Marchesi et al. 2018). The torus can be pictured as a compact reservoir of low-angular momentum dusty gaseous material, and/or part of a windy outflowing structure connected to the accretion disc (Hönig 2019, and references therein). Irrespective of its underlying nature, a torus around a SMBH significantly contributes to absorb UV light from the accretion disc and reprocess it in IR bands. As GAEA does not explicitly include the dynamical modelling of an accretion disc and a torus around the central SMBH, in what follows we include two torus models and also a combination of them: the model proposed by Wada (2015, Wada hereafter) [2] and the model proposed by Ramos Almeida & Ricci (2017, RA&R hereafter). The former model analytically connects the dependence of the torus size and thickness on AGN luminosity/accretion rate and SMBH mass, as detailed below, and assumes that in an AGN there is always enough circumnuclear material to feed a torus. The latter assumes that the column density increases for larger inclination angles, with maximum CTK column densities for the centre of the torus, with no explicit dependence on SMBH accretion rate or mass. We give further details below.

RA&R is based on the model where the fraction of the optical/UV and X-ray radiation processed by the torus and observed in the mid-infrared is proportional to its covering factor (Ricci et al. 2015, 2017b). Under this model, they assume that in the X-rays the covering factor of the gas and dust surrounding the SMBH can be estimated using a statistical argument and studying the absorption properties of large samples of AGN. Since a compact X-ray corona only gives information of that particular line-of-sight, a large sample study could provide further constrains in inclination angles and therefore other characteristics of the obscuring material. The intrinsic column density distribution of local hard X-ray selected AGN in the data of Ricci et al. (2015) shows an average roughly constant with luminosity covering factor (CF) of the obscuring material of 70%, implying a maximum opening angle of 45 degrees (CF = $\sin \theta$). Some other works, like Tanimoto et al. (2020, 2022); Ogawa et al. (2021); Yamada et al. (2021, 2023), propose torus opening angles somewhat larger between 60-90 degrees, corresponding to higher CF (CF ~ 90-100%) (see also Esparza-Arredondo et al. 2021, for CF>70% for Seyfer 2 galaxies in the X-ray and for Seyfer 1&2 in MIR). In order to match our targeted Compton-thick AGN observed fractions (U14; A19), in what follows we will adopt the same baseline structure of the RA&R model but with a slightly larger value of the covering factor, CF~93%, which is more representative of the latest observational results and, we found, simultaneously provides a better match to the data on the fractions of Compton-thick AGN. Following the RA&R layout, the CF is then subdivided into three areas, the low Compton-thin (CTN, $N_H \sim 10^{22} - 10^{23}$ cm$^{-2}$), high Compton-thin (CTN, $N_H \sim 10^{23} - 10^{24}$ cm$^{-2}$) and low Compton-thick (CTK, $N_H \sim 10^{24} - 10^{26}$ cm$^{-2}$), corresponding to 52-70 degree (CF ~ 93%), 27-52 degree (CF ~ 78%), and 0-27 degree (CF ~ 45%) angles, respectively (see Figure 2). We decide to modify the original RA&R parameters in order to produce higher covering factors, and assess the minimum CTK obscured fractions required from the torus to reproduce observations.

Wada suggests a kinematic model to describe the behaviour of the radiative feedback and the origin of the dependence of the obscured fractions on AGN luminosity. Their work proposes that the AGN produces a fountain of gas creating a radiation pressure on the dusty gas, with the accretion disc radiating most of its energy towards the direction of the rotational axis, not towards the plane of the disc. Besides, the radiative heating is isotropic affecting the surrounding gas through advection. They assume that the X-ray radiation from the AGN is spherically symmetric and heats the inner part of the thin gas disc, making it geometrically thick.

---

[2] We adopt the version of the code provided by Johannes Buchner: https://github.com/JohannesBuchner/agnviz

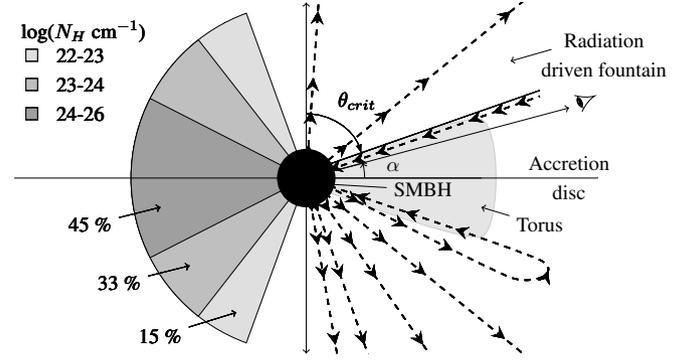

**Figure 2.** Diagram of the *fiducial* torus model where we implement both the radiation-driven outflows described by Wada, where $\theta_{\rm crit}$ is the critical angle of the radiation-accelerated gas (right side), and the RA&R constant limit angles (left side). $\alpha$ is the random line-of-sight angle. The diagram is separated in two for clarity.

The way we include the Wada torus model in our host galaxies is by assuming a random value of $\alpha$ line-of-sight, which compared with the torus critical angle $\theta_{\rm crit}$ provides us with a random column density, either $N_H \sim 10^{20} - 10^{24}$ cm$^{-2}$ (CTN obscuration) or $N_H \sim 10^{24} - 10^{26}$ cm$^{-2}$ (CTK obscuration). The selected values depend on the absence or presence of the torus, which is in turn determined by the line-of-sight $\alpha$ for both the torus, and the host disc components (see Fig. 1 and 2) and its relation with $\theta_{\rm crit}$, the critical angle of the radiation-accelerated gas (see Fig. 2). If $\alpha < 90 - \theta$, the line-of-sight lies within the torus, and the source is assumed to be CTK, otherwise it is considered a CTN AGN.

The critical angle $\theta_{\rm crit}$ originates from the balance between the radiation pressure on the gas and the gravitational potential of the SMBH. At any angle $\theta \leq \theta_{\rm crit}$, the radiation force is large enough to allow the dusty gas to escape, and therefore we expect gas outflows forming without the presence of the torus. In the region defined by $\theta > \theta_{\rm crit}$, the gas eventually falls back towards the equatorial plane, causing dusty gas to remain in the centre and forming a thick torus. The critical angle $\theta_{\rm crit}$ depends on both the SMBH mass $M_{\rm BH}$ and the bolometric luminosity $L_X$. More specifically, following Eq. 6 from Wada (2015), the critical angle is defined as

$$\cos \theta_{\rm crit} = \frac{GM_{\rm BH}}{r_0} \frac{16\pi c}{\kappa \gamma_{\rm dust} L_{\rm UV}} \left( \frac{1}{r_{\rm dust}} - \frac{2}{r_0} \right)^{-1} \quad (10)$$

where $c$ is the velocity of light, $\kappa = 10^3$ cm$^2$ g$^{-1}$ is the opacity of the dusty gas, $\gamma_{\rm dust} = 1/100$ is the dust-to-gas ratio, $r_0$ is the radius within which the X-ray heating is effective, $r_{\rm dust}$ is the dust sublimation radius, and $L_{\rm UV}$ is the UV luminosity of the AGN (Marconi et al. 2004). This UV luminosity is related to the X-ray luminosity of the AGN by $L_X = (1/2) \cdot L_{\rm UV} |\cos(\theta)|$ where $\theta$ is the angle from the rotational axis. All fixed values are the ones assumed in Wada. When using Eq 10, an increasing AGN power increases the angle $\theta_{\rm crit}$, reducing the chance of intersecting the torus. Note that it is assumed in this model that $r_0 > r_{\rm dust}$ which is always the case for $L_X < 10^{47}$ erg s$^{-2}$ if one defines $r_0$ as

$$r_0 = \left( \frac{3 L_X}{4\pi \Lambda_{\rm cool}} \right)^{1/3}, \quad (11)$$

where the radiative cooling with cooling rate $\Lambda_{\rm cool}$ is balanced by the X-ray heating rate $\rho_g^2 \Lambda_{\text{cool}} = 3L_X/(4\pi r_0^3)$, $\rho_g$ being the average gas density in the gas sphere with $r = r_0$. Wada also assumes that $r_d = 1.3(L_X/10^{46})^{1/2}$ pc (e.g., Lawrence 1991) and $\Lambda_{\text{cool}} = 10^{-22} n_H^{-2}$ erg cm$^3$ s$^{-2}$ ($n_H = \rho_g/m_p$, where $m_p$ is the proton mass).

**Table 1.** Summary of the different models studied in the paper. They are separated by its area of appliance (host galaxy, torus or both) and by model (no-BW/BW, RA&R/Wada/Fiducial, or a combination).

| Part | Model | Summary |
|---|---|---|
| Host | Fiducial host ($R_{d,\text{Wel}}$, no BW) | Model using Eq. 4. Column density calculated with a disc morphology using all the gas available. |
| | Fiducial host + BW ($R_{d,\text{Wel}}$, BW) | Model using Eq. 5. Column density calculated with a disc morphology from the gas left after the Blast Wave. |
| Torus | RA&R | Model where the X-ray raidation processed by the torus and observed in the MIR is proportional to its covering factor (left half of Figure 2). |
| | Wada | Model based on a radiative fountain (right half of Figure 2, Eq. 10). |
| | Fiducial torus | Combination of RA&R and Wada torus models. |
| Host + torus | Fiducial + noBW (Fiducial torus + host, noBW) | Combination of the fiducial torus model and the no BW host model. |
| | Fiducial + BW (Fiducial torus + host, BW) | Combination of the fiducial torus model and the BW host model. |

A combination of the two models is also studied, and it is labelled in the following plots as the *fiducial* torus model. In this combined model, we first calculate the critical angle from the Wada model, and we distinguish CTN from CTK depending on the angle $\alpha$ with respect to $\theta_{\text{crit}}$. More specifically, if $\alpha > 90 - \theta_{\text{crit}}$, then the AGN will be CTN with $N_H < 10^{22}$ cm$^{-2}$, if $\alpha < 90 - \theta_{\text{crit}}$, then three possibilities can arise following RA&R: 1) if $\alpha < 27$ deg then CTK, if $27 < \alpha < 52$ deg high CTN, $52 < \alpha < 70$ deg low CTN.

We include these dust obscured torus models in each galaxy in post-processing in the GAEA catalogues. A summary of all models can be found in Table 1.

## 3 RESULTS

In this section we compare the predicted mean $N_H$ column densities and obscured AGN fractions as a function of AGN X-ray luminosity, with data from Ueda et al. (2014), Buchner et al. (2015), and Ananna et al. (2019) (U14; B15; A19, hereafter), which are among the most complete compilations in terms of AGN luminosity and redshift coverage, including data from deep surveys from observatories such as *Swift*/BAT, ASCA, *XMM-Newton*, *Chandra*, *ROSAT*, or *AEGIS*. The major differences between U14, B15, and A19 are described and deeply studied in A19, Sections 3.1 and 3.2, where they focus on the different methods to calculate the X-ray luminosity function. Specifically, we use two forms from A19, one which closely follows the analytic formula by U14 with updated parameters, and a new one derived from Machine Learning algorithms, which we label as A19-ML throughout. It is interesting to note that the two A19 prescriptions are fits to the same data sets but with different prescriptions, and provide very different results, as we show in the next Figures.

We need to take into account that the obscured values from U14 are extrapolations for column densities log $N_H > 24$ [cm$^{-2}$] since they do not have CTK AGN in their samples. U14 use a parametric model to fit the X-ray luminosity function (XLF), but could not directly constrain the CTK fraction, which is derived from matching the X-ray background (XRB) with some assumptions on the spectra of AGN. U14 do not assume any specific constraints for the CTK, but each bin of CTK is the same as the CTN fraction. They also use data that do not have any galaxy detected over log $N_H \sim 25$ [cm$^{-2}$] (except in infrared, but not resolvable in X-ray, for an updated analysis of U14 see also Yamada et al. 2021). While B15 have galaxies with CTK obscuration, A19's analysis extrapolates for log $N_H > 25$ [cm$^{-2}$]. The aim of A19 was to calculate new fractions where no assumptions on the CTK AGN fraction were given. Also, updating some B15 constraints, according to A19, does not help to match the current data. B15 use two different approaches: one where they have high CTK at high luminosity, overestimating the XLF by 3 times (see appendixes of A19), which produces around 55%-65% CTK, and another one where they use a constant slope prior which generates a lower CTK obscuration, around 20%. In this paper we use the 10%–90% quantiles of the posterior samples from both models as limits of the data. The major problem is that the uncertainty of the spectra distributions among the data available is not consistent with each other, and some of them never produce the cosmic X-ray background, which affects the conversion between number counts and flux. In this work we do not address the origin of these discrepancies but rather the source is use the total of the available observational results assumed to be CTK to bracket the current empirical constraints on the fraction of obscured AGN as a function of luminosity and redshift. Despite U14 and A19 not attempting to directly identify CTK AGN by, e.g., spectral fitting (see B15), we use those results as observational constraints as broad guidance for the fraction of CTK AGN required to match the normalization and shape of the X-ray Backgrounds (XRB). Our results therefore on, e.g., the need for a torus-like component to generate more CTK sources, rest on the future validation of the current observational constraints. We acknowledge, for example, other interesting works such as Akylas et al. (2012, see also Treister et al. 2009), who put forward models able to fit the XRB without any CTK AGN, but by modifying the X-ray spectrum of AGN, or Georgakakis et al. (2017), who suggest lower fraction of CTN AGN with $L_X > 10^{44}$ erg/s using the wide-area XMM-XXL survey. We conclude this discussion on the present-day constrains on obscured AGN noticing that a recent work, Laloux et al. (2023), combined X-ray spectral analysis with SED fitting to constrain the obscuration of a large sample of AGN. Overall, their results point to relatively large CTK fraction consistent, if not higher, than, those calibrated by U14. However, their error bars are still large to derive any firm conclusion on the true underlying fraction of CTK AGN.

### 3.1 The role of galaxy size and AGN feedback in shaping the obscured AGN fraction with $L_X$

In what follows, when discussing the dependence of AGN obscured fractions on X-ray luminosity, we focus on the mean redshift of $z = 2.4$, around the peak of AGN emissivity with available observational constraints. We will then show the fiducial model against data in other bins of redshift. In the left panel of Figure 3, we provide a comparison of the predicted mean column densities from our models as a function of X-ray luminosity compared with the mean empirical column densities extracted from the average

$$\langle \log N_H \rangle = \int f(\log N_H | L_X) \cdot \log N_H \cdot d \log N_H \quad (12)$$

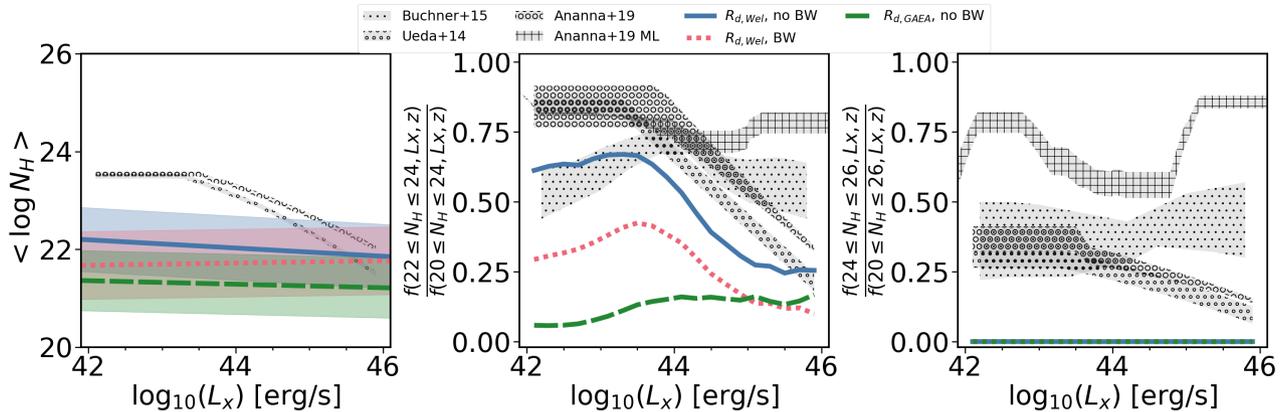

**Figure 3.** Mean column density distribution and obscured fractions at z = 2.4 depending on the X-ray luminosity. Green dashed lines corresponds to the model using the (gas) scale length from GAEA and without BW. Blue solid and red dotted lines are calculated with the scale length fit from van der Wel et al. (2014) no-BW and BW models respectively. All three lines assume $R_{d,gas} = 0.3 \cdot R_{d,\star}$. **Left panel:** Column density distributions predicted by each model, as labelled. Lines correspond to the mean values of the column density at fixed AGN X-ray luminosity, and the coloured areas mark the predicted $\sigma$ region around the mean. **Middle panel:** CTN obscured fractions. Grey areas correspond to the observations by U14, B15, and A19 as labelled. **Right panel:** CTK obscured fractions. Observations with the same format as in the middle panel.

where $f(\log N_H | L_X)$ is the conditional column density distribution derived by U14 and A19. Our results clearly highlight the importance of the correct recovery of the gas scale radii as a function of galaxy stellar mass, as seen from the significant difference between the AGN obscured fractions using GAEA disc radii and van der Wel et al. (2014) relation. In our reference model we assume gas disc thickness as $h = R_d/8$ following Nath Patra (2020). We will show the effect on our prediction of a different assumption for $h$ in the Appendix A. As robust and extensive measurements of the gas sizes are only available for sporadic samples (see Nelson et al. 2016; Puglisi et al. 2019, for comparison between ionised gas or cold gas and stellar disc radii in a statistical sample of $z \sim 1.5$ galaxies), in what follows we assume the $R_{d,gas} = N \cdot R_{d,\star}$, with $R_{d,\star} = 1.68 R_{eff}$, and $N = 0.3$, inspired by the recent ALMA/sub-millimetre observations by Puglisi et al. (2019) suggesting that on average the gas disc radius is about 1/3 of the stellar component, which is at variance with previous works that assumed $R_{d,\star} = R_{d,gas}$ (e.g., Tamburro et al. 2008; Leroy et al. 2008; Swinbank et al. 2017; Gilli et al. 2022; Liao et al. 2023). We note that assuming large gas scale lengths comparable to the stellar disc ones would induce too low CTN fractions, as shown in Figure 4, when adopting exponential profiles. We first use the gas disc sizes directly predicted by the GAEA model, and we obtain a weak positive dependence of $N_H$ versus AGN luminosity, which is at odds with observational constraints. Puglisi et al. (2019) results are complete on the main sequence only above $10^{11} M_{sun}$, but there are no results for lower masses. In order to explore the effect of the dependence of disc sizes on stellar mass, we also use the fitting formulae from van der Wel et al. (2014, Eq. 3 and Table 1). Using this empirical model, the situation clearly improves and the observed dependence of $N_H$ with X-ray luminosity is recovered, although its slope is still shallower than in the observed data. It is important to keep in mind that van der Wel et al. (2014) measure half-light radii of the stellar component. It is also worth noticing that the GAEA model predicts a disc size versus stellar mass relation which is consistent with van der Wel et al. (2014) data (Zoldan et al. 2019), but with a slightly shallower slope. Our results thus highlight the need for a relatively steep disc size vs stellar mass relation in order to recover the trend of obscuration with bolometric luminosity. We explore the impact of varying the ratio $N = R_{d,gas}/R_{d,\star}$ within reasonable values in Figure 4, which shows that only models with $N \lesssim 0.3$ can generate a fraction of CTN AGN broadly consistent with current data (solid, blue and dotted, green lines). The effect of varying the gas disc thickness instead is marginal and definitely negligible with respect to the impact of the BW and/or the choice of gas disc sizes, as discussed in Appendix A.

The middle panel of Figure 3 compares the predicted fraction of obscured CTN AGN, with the data by U14, B15, and both of the A19 models, as labelled. The *fiducial* model (without BW, assuming the stellar mass dependence of the $R_{d,\star}$ from van der Wel et al. (2014), and $R_{d,gas} = 0.3 R_{d,\star}$) presents a decreasing trend in the obscured fraction with increasing X-ray luminosity, which is also present when including the BW model. This decreasing trend, which is aligned with observations (e.g., Gilli et al. 2007; Hasinger 2008; Ueda et al. 2014; Buchner et al. 2015; Ananna et al. 2019), is mainly induced by the lower luminosity AGN, which tend to have a relatively higher fraction of high column densities being generally hosted in lower mass and more compact galaxies. As seen in Eq. (4), in fact, at fixed line-of-sight angle, a smaller $R_{d,gas}$ would increase the upper end of the integral and thus the corresponding $N_H$.

When including the BW in Figure 3, our predicted fractions drop by ~30% at low luminosities and ~15% at higher luminosities. Although the impact of the BW is somewhat degenerate with the exact choices of gas fractions and/or shape of the still poorly constrained $R_{d,gas} - M_\star$ relation, it is still relevant to highlight two effects of the BW model. First, with all other parameters kept fixed, the BW model makes it usually harder for galaxy scale obscuration to make a significant contribution to the fraction of obscured AGN, due to some gas being removed from the galaxy when the BW is bigger than the extension of the gas disc. Second, at least within the remit of the Menci et al. (2019) model, the BW is not the cause behind the drop in the fraction of obscured AGN with luminosity, a trend which in our model is instead mostly driven by the (positive) correlation between $R_{d,gas}$ and $M_\star$.

The right panel of Figure 3 shows that the galaxy-scale obscuration, irrespective of the specific parameters adopted in input, falls drastically short in producing any CTK AGN at any X-ray luminosity, at least in the case of an exponential gas density profile, suggesting that something in the current model is still missing.

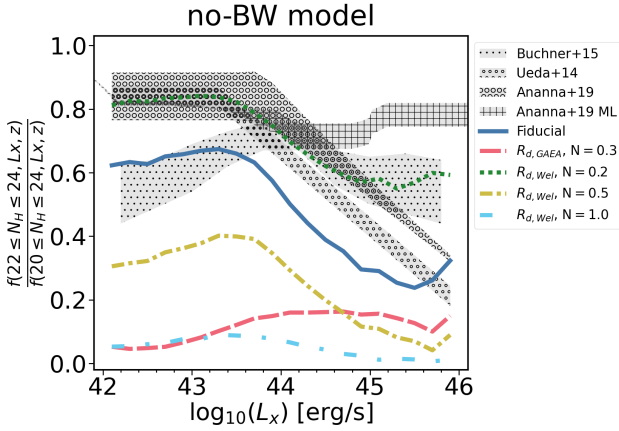

**Figure 4.** Predicted fraction of CTN obscuration as a function of X-ray luminosity and without a BW for different values of the relation between gas disc radius $R_{d,gas}$ and stellar disc radius $R_{d,\star}$, $R_{d,gas} = N \cdot R_{d,\star}$, as labelled, with a fixed thickness of $h = R_d/8$ at redshift $z = 2.4$. We explore $N = 0.2$, $N = 0.3$, $N = 0.5$, and $N = 1.0$. Fiducial model refers to the model using van der Wel et al. (2014) fit and $N = 0.3$ (see text). The observational data is shown as in Figure 3.

To further clarify the importance of the dependence between gas disc sizes and galaxy stellar mass, in Figure 4 we plot the predicted fraction of CTN obscuration for four different values of the variable $N$ of the $R_{d,gas} = N \cdot R_{d,star}$ relation, where the fiducial model (solid blue line) is $N = 0.3$ as previously defined, and the other three lines are for different choices of $N$, as labelled. We also include a model with the original GAEA $R_d$ for comparison (cyan long dot-dashed line). The smaller the $N$ value, the larger the overall implied mean column densities. The value $N = 0.3$ has been chosen following the results of Puglisi et al. (2019), which is the average ratio between the stellar and sub-mm radius in sub-mm compact galaxies. Lower $N$ generates more compact and obscured galaxies, to the point where we can reproduce the full fraction of CTN as measured by U14 (see N=0.2, dotted green line for U14; A19 at lower luminosities and B15; A19, -updated Ueda version- at higher luminosities). This trend indicates that, with sufficiently compact galaxies, we can reproduce the CTN obscured fractions without the need for any other obscuration component. However, due to the lack of extensive measurements of the molecular gas disc size in statistical samples of main-sequence galaxies, we cannot confirm (nor reject) that all galaxies at high redshift present gas scale lengths below 0.7 kpc, which are the values obtained when assuming $N = 0.2$. In this paper we choose a value of $N = 0.3$ in order to be conservative, and in line with some of the latests observations (Puglisi et al. (2019), see also Elbaz et al. 2018; Franco et al. 2020; Puglisi et al. 2021; Gómez-Guijarro et al. 2022). Meanwhile, we have confirmed that any combination of input parameters explored in this paper can hardly generate any significant number of CTK AGN, although a few more could be formed when switching to a Sérsic gas density profile, as further discussed below.

The analysis of Figures 3 and 4 has been carried out under two major assumptions: 1) the gas fractions do not evolve significantly during the life span of the AGN, and 2) the X-ray luminosity associated to $N_H$ for each source is the peak luminosity within the AGN light curve. The former assumption might be extreme as gas fractions decay in time due to gas consumption via star formation and, as predicted by many galaxy evolution models, via AGN feedback which can both heat and expel gas (e.g., Granato et al. 2004; Hopkins et al. 2006; Croton et al. 2006). The second assumption is a somewhat natural choice in GAEA as AGN light curves tend to be quite narrow due to an emission bulk highly concentrated in time, a combination of large initial SMBH masses (most of the AGN in GAEA at $z < 3.3$ are re-activations), and a rapid fading phase for less luminous objects (the large majority of events). In addition, following F20, the peak of the light curve is shorter because the bolometric luminosities significantly drop when they enter the radiatively inefficient mode, below 10% Eddington luminosity. In order to check the robustness of our conclusions against the above assumptions, we develop several additional models where either a) we associate an X-ray luminosity randomly chosen within the AGN light curve to the column density, b) we deploy a column density which decreases exponentially over time, or c) we assume a more extended input AGN light curve. We report the results of our new additional models in Appendix B, where we show that, in all cases, our main results are similar to the ones obtained in our fiducial model.

As a final check, we compare our fiducial host model against U14 and A19 for the obscured fraction distributions in the column density plane in Figure 5. We note that the column density distributions as a function of X-ray luminosity from A19 were recently confirmed in the mid-infrared (MIR) at $z \leq 0.8$ by Carroll et al. (2023). We immediately note that, as expected from our previous findings, our reference models fall severely short in matching the fraction of CTK AGN when only the obscuration from the host galaxy is included. In addition, Figure 5 also reveals that, although our fiducial model (without a torus) can predict an overall inverse dependence of obscured fraction with increasing X-ray luminosity similarly to what observed in the data (Figure 4), it still struggles in fully reproducing the breakdown of CTN AGN at fixed X-ray luminosity. The fiducial model generates similar fractions of $N_H \sim 10^{22} - 10^{23}$ cm$^{-2}$ as in the data, but less AGN with $N_H \sim 10^{20}$ cm$^{-2}$ and significantly more AGN with $N_H \sim 10^{21}$ cm$^{-2}$. We will see below that including a torus in our fiducial model provides an improved match to the data on the $N_H$ distribution at fixed X-ray luminosity.

### 3.2 The contribution of the torus to the AGN obscured fraction

So far, we have been considering only the contribution to the $N_H$ column density of the large-scale distribution of gas in the host galaxies. We now proceed with the inclusion of the torus as an independent source of AGN obscuration. Figure 6 shows the three models described in Section 2: Wada, RA&R and the *fiducial* torus model, a combination of the other two with no contribution to the obscuration from the host galaxy.

The implementation of the Wada torus model (red, dotted lines of Figure 6) in our mock galaxy catalogue produces, using their suggested parameters, a significant fraction of CTK AGN of ~85% at low luminosity, with a steep decrease to ~25% at brighter luminosities (right panel of Figure 6). This trend is mostly a consequence of the dependencies of the torus radius (the radius within which the X-ray heating is effective) and $\theta_{crit}$ on luminosity, with the former increasing and the latter decreasing with increasing luminosity (see Section 2.3). Both variables are contributing by lowering the probability for the central SMBH to be obscured along any random line-of-sight, especially in the more luminous AGN. The torus also significantly contributes to the obscuration of AGN in the CTN regime (middle panel). The average value of 50% of CTN obscuration across all luminosities (middle panel of Figure 6) naturally arises from our adopted assumption (see Section 2.3) of assigning a column density to all non-CTK obscuration uniformly distributed between $20 < \log(N_H/cm^{-2}) < 24$.

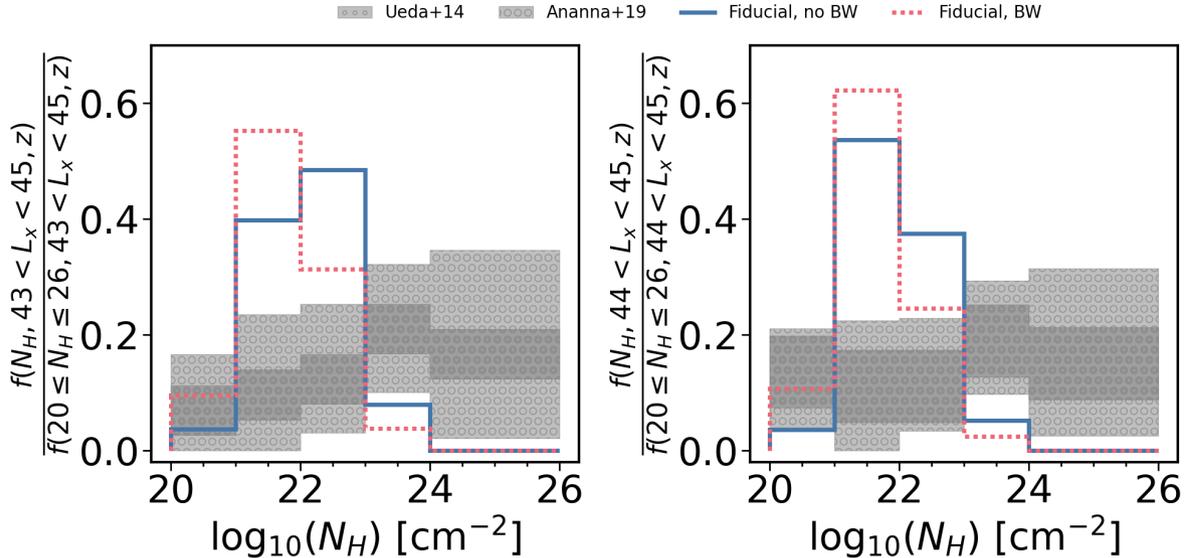

**Figure 5.** Obscured fractions of galaxies between $L_{\rm X} = 10^{43}$ erg/s and $L_{\rm X} = 10^{45}$ erg/s (left panel) and between $L_{\rm X} = 10^{44}$ erg/s and $L_{\rm X} = 10^{45}$ erg/s (right panel) as a function of the column density for the fiducial host model with (dotted red line) and without (blue solid line) BW at redshift $z = 2.4$. Observational data correspond to U14 and A19, as labelled.

Figure 6 also includes the RA&R model (long-dashed, green lines), which relies on constant limit values for the $\alpha$ angle depending on the column density and the line-of-sight, creating constant fractions of obscured AGN for both CTN and CTK. In the case of the CTN, the model presents a mean value around 67%, while the CTK predicted fraction is ~30%, with negligible dependence on AGN luminosity, as expected. The predicted fractions of CTN and CTK AGN from this torus model alone are already significant enough to be comparable to the observations of B15 for both CTN and CTK.

Our *fiducial* torus model (solid, blue lines in Figure 6) includes the luminosity dependent features of the Wada torus model, as well as the angle dependency of the $N_H$ distribution from RA&R. This fiducial model, in line with the RA&R model, naturally predicts a ~30% fraction of CTK at low luminosities (right panel), reflecting the assumed value from RA&R model that sources below 27 deg are CTK, but gradually decreasing to a few percent at bright luminosities due the (negative) luminosity dependence of the opening angle. A similar trend is observed in CTN (middle panel), showing a value at faint luminosities close to the one predicted by the RA&R model, and then gradually decreasing at higher luminosities.

In Figure 7, we sum the predicted large-scale obscuration from the host galaxy gas with the small-scale obscuration from the torus for different redshifts. Our fiducial torus and host model, which is the combination of our preferred models from Figure 3 and Figure 6, is reported here, with and without the inclusion of the BW (red, dotted and blue, solid lines, respectively). Our reference model provides a good match to the U14 data at least at $z > 2$. Overall, the fraction of CTK is roughly constant across cosmic times, and slightly decreasing at lower redshifts and at luminosities below $L_{\rm x} \sim 10^{44}$ erg/s. The fraction of CTN AGN is also roughly constant at $z \gtrsim 2$, but then steadily decreasing in normalization with cosmic time especially at lower luminosities. The progressively increasing drop in the fractions of low luminosity obscured AGN at $z < 1$ is mostly driven by the increasing number of Radio-mode sources in the model. In Figure 7, the yellow dot-dashed line shows a realization where we compute the bolometric luminosity using QSO-mode accretion only: this implies that we remove from the estimate all sources powered by Radio-mode accretion, that are not obscured by the host by construction (since their gas content is almost zero), which results in an increase of the obscured fractions. Radio-mode has a marked effect only in the $z \sim 0$ panel, due to the overall decline of the AGN space density and the increase of massive and gas-poor galaxies which are not largely represented in the sample of obscured X-ray AGN. In this redshift range, neglecting the Radio-mode accretion in the luminosity calculation increases the CTN and CTK AGN obscured fractions at low luminosity, bringing them in better agreement with the available constraints. This is mainly due to the fact that by removing Radio-mode dominated sources we are preferentially removing model galaxies that are expected to be unobscured in our modelling. Indeed, Radio-mode dominated sources are mostly massive galaxies, that, by construction, have a negligible gas content and low bolometric luminosities. As expected, the impact of removing the Radio-mode channel on our predictions strongly decreases at increasing redshift and is completely marginal at the redshift of interest for this paper.

For completeness, we compare our fiducial model against U14 and A19 for the obscured fraction distributions in the column density plane in Figure 8. We note that the inclusion of the fiducial torus component improves the match to observations compared to a model inclusive of only the obscuration from the host galaxy (Figure 5). In the left panel we show the fractions of all galaxies within X-ray luminosities between $10^{43}$–$10^{45}$ erg/s. For column densities between $N_{\rm H} \sim 10^{20}$–$10^{21}$ cm$^{-2}$, we still predict a relative deficit of obscured sources. However, the model tends to better align with the data at larger $N_{\rm H}$ column densities, although the uncertainties in the current available data are still significant. The full model host galaxy+torus tends to smooth out the sharp peak observed before in Figure 5, in better, albeit not perfect, agreement with the data. We note that the fraction of CTK AGN we predict from our reference model is never too large, roughly consistent with the one inferred by U14 and A19 from fits to the X-ray background, but never beyond the ~10-15%

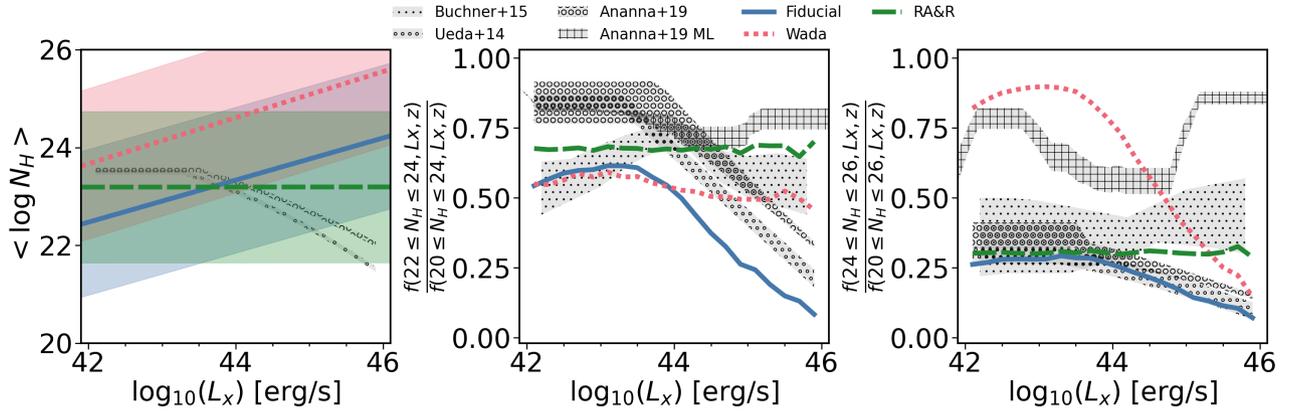

**Figure 6.** Column density distribution and obscured fractions depending on the X-ray luminosity for different types of torus models at redshift $z = 2.4$. Fiducial model (blue solid line) refers to a combination of Wada and RA&R torus models (see text), red dotted line shows the Wada torus model, and green dashed line is the RA&R model. Observations shown as in Figure 3. **Left panel:** Mean column density distribution along with its $\sigma$ (coloured area). **Middle panel:** CTN obscured fractions. **Right panel:** CTK obscured fractions.

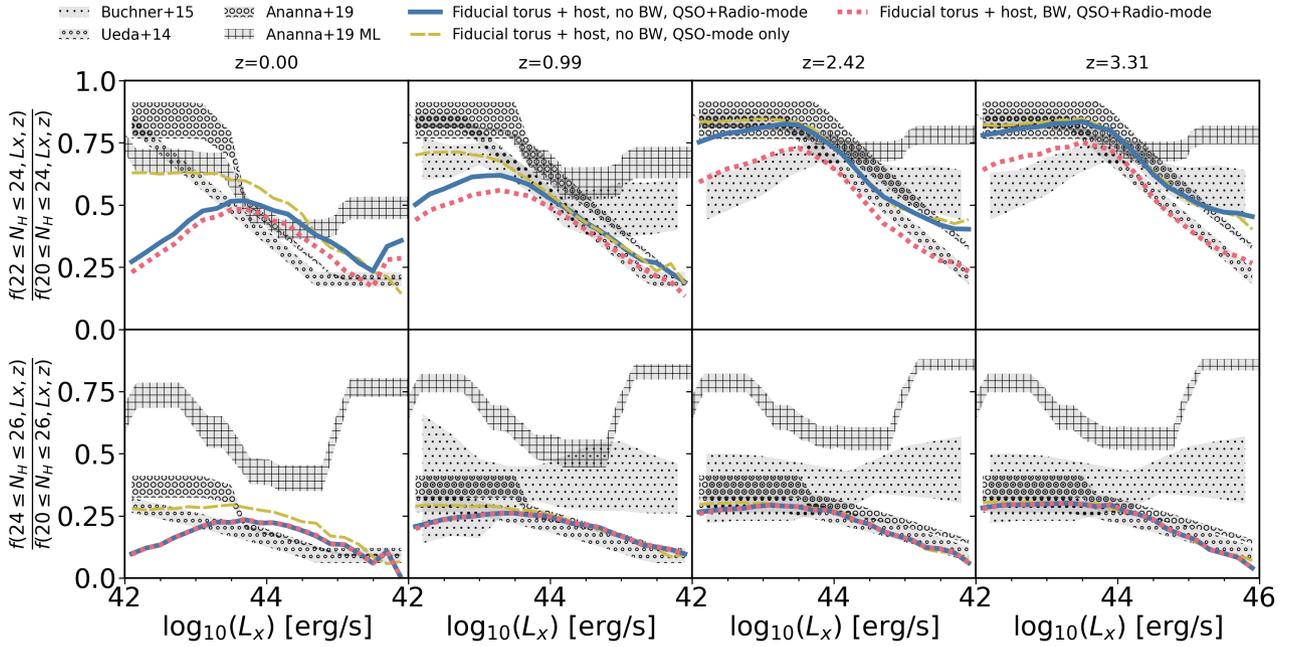

**Figure 7.** Compton-thin and thick obscured AGN fractions depending on the X-ray luminosity for galaxies including BW (dashed red) or without the BW (solid blue line) and the fiducial torus model. Each panel corresponds with a different redshift as labelled. Observations shown as in Figure 3.

limit. The inclusion of the AGN BW has a minimal impact on our predicted $N_H$ distributions.

## 4 DISCUSSION

By modelling the small- and large-scale obscuration of AGN on a galaxy-by-galaxy basis, we have been able to pin down, in the context of an Orientation model, the main parameters driving AGN obscuration (namely galaxy structure and the torus component). Here we discuss the robustness of these results starting from our assumptions, compare with other related works in the literature, and connect with Evolutionary models.

Our reference model for this study is the state-of-the-art semi-analytic model GAEA, which provides a self-consistent baseline population of galaxies, with their central SMBHs and accretion rate distributions consistent with the total AGN luminosity function (see F20). The aim of this Section is to probe the impact of our results on some underlying assumptions and also input parameters. In an empirical/data-driven fashion, we thus change in turn some of these main input parameters. We have already seen that although GAEA correctly predicts the increase of the mean galaxy size with increasing galaxy stellar mass, only when assuming the steeper empirical relation from van der Wel et al. (2014) we obtain the right trend of $N_H$ with $L_X$. However, an $R_{d,\star} = R_{d,gas}$ relation is insufficient to reproduce the necessary CTN sources to reproduce observations (Figure 4). We thus need a more compact gas component, as suggested by recent results from ALMA by Puglisi et al. (2019, 2021). Further observational constraints in the cold gas mass disc radii are needed in order to test our results, such as measurements of molecular gas disc sizes, and/or larger galaxy samples with AGN detections. We also study the effect of varying, within observational constraints, the gas disc scale height $h$ (see Appendix A). We find that the gas scale height $h$ plays a minor role in the overall results when compared to other variables like the BW.

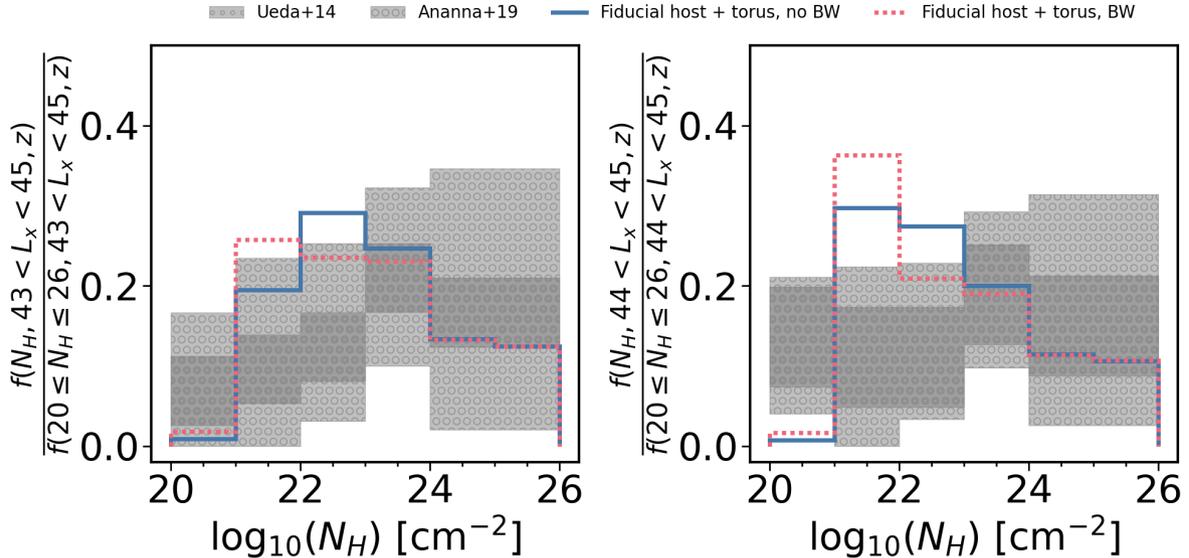

**Figure 8.** Obscured fractions of galaxies between $L_X = 10^{43}$ erg/s and $L_X = 10^{45}$ erg/s (left panel) and between $L_X = 10^{44}$ er/s and $L_X = 10^{45}$ erg/s (right panel) as function of the column density for the fiducial galaxy model (host and torus) with (dotted red line) and without (blue solid line) BW at redshift $z = 2.4$. Observational data correspond with U14 and A19, as labelled.

Additionally, the modelling of an AGN-driven BW feedback capable of removing significant portions of the cold gas mass from the inner regions, affects somewhat the normalization of the fraction of obscured AGN, but not its luminosity dependence. The only dependence of the BW with the luminosity comes from the opening angle calculation, which will produce all possible angle values between $\log_{10} L_X = 40 - 44$ [erg/s] to then exponentially saturate at 90 degrees between $\log_{10} L_X = 44 - 46$ [erg/s]. Therefore, a luminosity dependence of the BW will only slightly affect higher luminosities, as seen in Figure 7. The two variables of gas disc sizes and the impact of a BW appear therefore somewhat degenerate, as increasing the latter requires steepening the former. More robust constraints on the $R_{d,gas} - M_\star$ and/or the presence of BWs in AGN will help in further guiding the models.

Other assumptions seem to play a minor role in setting the obscuration levels in AGN. For example, we checked the effect of adopting the empirical gas fractions as function of stellar mass and SFR by Santini et al. (2014), which yields similar results to our reference model based on the GAEA gas fractions (see Appendix C).

By applying our *fiducial torus* model to our galaxies, we are including a luminosity dependency on the obscured fractions, as seen by U14 and A19, which ultimately originates from the radiative fountain of Wada (Figure 2). However, the results obtained by B15 or A19 ML suggest a less dependent obscured fraction with luminosity, more in line with RA&R model, which alone falls short in reproducing the observations despite our increase in the covering factor following new results (e.g., Tanimoto et al. 2019, 2020, 2022; Ogawa et al. 2021; Yamada et al. 2021, 2023). Including a luminosity dependence in this model could be achieved by including a dependency on the Eddington rate instead of a constant value of the torus critical angle for all galaxies (Ricci et al. 2017a; Yamada et al. 2021; Ogawa et al. 2021).

When exploring the light curve of the AGN (see Appendix B), we show that associating to each galaxy's column density the peak X-ray luminosity or a random one within the light curve yields similar results, mainly due to the relatively briefly peaking light curve predicted in GAEA. One might wonder the consequence of extending the light curve and thus broadening the choice of X-ray luminosity to map to each $N_H$. We carried out this exercise in Appendix B, finding similar results to our reference model. It is interesting to note that the adoption of extended light curves also allows for exploring the impact of evolutionary patterns on the obscured fractions (e.g., Sanders et al. 1989; Granato et al. 2004, 2006; Lapi et al. 2006; Hopkins et al. 2010). For example, by assuming that obscured/unobscured AGN sources are preferentially those observed in their pre-peak/post-peak phase, it would be possible to test whether traditional evolutionary models could reproduce the latest observations on the fractions as a function of luminosity and redshift. A preliminary investigation shows that forcing obscured sources $N_H > 10^{22}$ cm$^{-2}$ to only appear in the pre-peak phase does not change our results. Another interesting test involving the torus appearance only during the pre-peak phase of the AGN light curve is also included in Appendix B. This test confirms that the evolution of the torus can affect the predicted fractions of obscured AGN, which may even fall below current observational constraints in some instances. Our results suggest that, despite the lifetime of the AGN torus being somewhat degenerate with the type of specific torus model adopted in the context of our model, to produce CTK obscuration in our models the presence of an inner torus appears to be an essential and ubiquitous feature. This is consistent with the analytic calculation carried out by Buchner & Bauer (2017, Appendix B), which demonstrated that CTK column densities cannot be achieved by accumulating the galaxy gas over several central kpc.

Our current work, strictly based on exponential gas disc profiles, suggests that, at least at $z \lesssim 3$, galaxy-scale obscuration may not be sufficient to account for the significant fraction of CTK AGN, and may even fall short in reproducing all CTN AGN. Nevertheless, some care is needed in extrapolating this conclusion at higher $z$. In fact, in many high-z star-forming galaxies (harbouring a growing central BH), most of the host galaxy obscuration is not associated to an extended gaseous HI disk, but rather to a roughly spherical and compact (about 1 kpc) central region, rich in molecular gas and dust, where most of the star-formation is taking place (e.g., Knapen et al. 2006; Chen et al. 2016; Molina et al. 2023). There, the equivalent gas column densities may be extremely high, to provide an obscuration comparable, or even heavier than from the nuclear torus (cf. Gilli et al. 2022). We reserve the modelling of these highly obscured systems in a forthcoming paper, where we will explore the full impact of evolutionary models on AGN obscuration. Nonetheless, we have carried out a preliminary test using a 3D Sérsic density profile (Prugniel & Simien 1997) assuming $R_e = R_d/1.68$. This calculation suggests the possibility of fully reproducing the CTK fractions with only the host galaxy component with Sersic index $n = 2 - 3$, although saturating the CTN fractions to 100%. Further targeted observational and theoretical work is needed to verify and confirm these results.

Our best model is in agreement with the recent predictions by Gilli et al. (2022) who propose that the total covering factor from the interstellar medium within galaxies is not sufficient to produce CTK obscuration at $z \lesssim 3$. However, its contribution to obscuration can drastically increase at higher redshifts due to an overall strong increase in the gas cloud surface density in the host galaxies.

## 5 CONCLUSIONS

The source of obscuration in AGN is still highly uncertain, as it can arise from the large-scale obscuration of the galaxy, and/or from an inner dusty torus component around the central SMBH. In this paper, we have modelled from first principles, in the framework of a comprehensive semi-analytic model, the origin of obscuration in AGN in the context of pure orientation models with a first incursion into some evolutionary components that might play a key role in the obscuration definition. Our main results can be summarized as follows:

- On the strict assumption of an exponential cold gas density profile, we find that the fraction of CTN obscuration contributed by only the large-scale galaxy obscuration is not enough to reproduce the current observational constraints, unless we assume very compact galaxies while also having the gas disc scale length increasing with stellar mass, as measured by, e.g., van der Wel et al. (2014) for the stellar discs (Figures 3, and 4).
- The inclusion of a physically-motivated, AGN-driven shock BW reduces the gas fractions and thus the overall $N_H$ column densities. However, the BW is not the main driver behind the drop of the obscured CTN fractions with X-ray luminosity, which is mostly driven by the gas disc sizes increasing with stellar mass (Figure 4). Our results point to the morphology of the cold gas component as the main driver shaping the properties of the obscuration of AGN, at least in CTN sources.
- Irrespective of the exact parameters and model assumptions, the large-scale gas distributions fall short in reproducing any significant fraction of CTK obscuration (Figures 3 and 5), at least at $z \leq 3.3$, and when adopting a strictly exponential profile for the cold gas component.
- The inclusion of a dusty torus with opening angle depending on both AGN luminosity and BH mass as in Wada (2015) with the $N_H$ limits discussed by Ramos Almeida & Ricci (2017) (adopted as a fiducial torus model in this work), nicely matches the full distribution of CTK obscuration as a function of X-ray luminosity, and also contributes to the fraction of CTN AGN. The full fiducial model also broadly, albeit not perfectly, aligns with the AGN $N_H$ distributions at fixed X-ray luminosity.
- Within the remit of the model explored here, the presence of an inner torus appears to be an essential and ubiquitous contributor to AGN obscuration, especially for the more luminous CTN and most of the CTK sources.
- A time-dependent torus model disappearing in the post-peak phase might be able to reproduce the CTN and CTK obscuration, but it heavily depends on how the torus and the AGN light curve are modelled (see Appendix B).

Our core results are robust against variations in input AGN light curves, galaxy gas masses and disc morphology. Our work has highlighted the key importance of a combined contribution of small and large-scale obscuration to provide a full census of AGN at $z < 3.3$. On the other hand, some relevant points remain to be investigated, namely the contribution of mergers and/or dust-enshrouded/highly star-forming galaxies in controlling the demography of obscured AGN, especially at high redshift, as well as the caveats mentioned in the discussion. We aim to address these, and other Evolutionary features contributing to AGN obscuration, in a forthcoming paper.

## ACKNOWLEDGEMENTS

We thank the referee for their useful comments that significantly improved the presentation of results and comparison with the data. This study has been carried within BiD4BESt [3], an Innovative Training Network (ITN) providing doctoral training in the formation of supermassive black holes in a cosmological context. BiD4BESt has received funding from the European Union's Horizon 2020 research and innovation programme under the Marie Skłodowska-Curie grant agreement No 860744 (grant coordinator F. Shankar). FF, and NM acknowledge support from PRIN MIUR project 'Black Hole winds and the Baryon Life Cycle of Galaxies: the stone-guest at the galaxy evolution supper', contract 2017-PH3WAT. AVAT thanks Trieste Observatory (Italy) for hospitality during part of the development of this work. AL is partly supported by the PRIN MIUR 2017 prot. 20173ML3WW 002 'Opening the ALMA window on the cosmic evolution of gas, stars, and massive black holes'. AVAT also acknowledges Tonima Ananna and Claudio Ricci for helpful discusisons. AP acknowledges partial support by STFC through grants ST/T000244/1 and ST/P000541/1. CRA acknowledges support from the project "Feeding and feedback in active galaxies", with reference PID2019-106027GB-C42, funded by MICINN-AEI/10.13039/501100011033. MV is supported by the Alexander von Humboldt Stiftung and the Carl Friedrich von Siemens Stiftung. MV also acknowledges support from the Excellence Cluster ORIGINS, which is funded by the Deutsche Forschungsgemeinschaft (DFG, German Research Foundation) under Germany's Excellence Strategy - EXC-2094 - 390783311. VA acknowledges support from INAF-PRIN 1.05.01.85.08. SB acknowledges partial support from the project PID2021-124243NB-C21 funded by the Spanish Ministry of Science and Innovation.

---

[3] More information about BiD4BESt and the Innovative Training Network can be found in https://www.bid4best.org/.

**DATA AVAILABILITY**

An introduction to GAEA, a list of our recent work, as well as datafile containing published model predictions, can be found at https://sites.google.com/inaf.it/gaea/home. The analysis carried out in this work as well as the plotted results will be available upon request until a free access database is released (which will be found in https://github.com/AVAlonso).

## APPENDIX A: DEPENDENCE ON GAS DISC SCALE THICKNESS

In this work we have assumed that a reasonable definition for the disc thickness is $h = R_d/8$, as suggested by, e.g., Nath Patra (2020). Here we explore the impact on our results when adopting a different definition. The one by Ojha (2001), who proposes $h = R_d/15$, leads to thinner discs for all galaxies. Other works also use $h = 0.15 \cdot R_d \sim R_d/6$ as fixed value (e.g., see Gilli et al. 2022, and references therein), which leads to thicker discs. We assume constant disc thickness throughout the redshifts studied, as observed by Hamilton-Campos et al. (2023) in galaxies $z > 1$.

In Figure A1 we show the CTN obscured fractions without BW and for the three different disc thickness definitions. The exact value of disc thickness $h$, when chosen within the observational range, does not significantly alter the overall shape of the predicted CTN fractions, except for a luminosity-dependent increase of around $10-20\%$ at all X-ray luminosities. The disc thickness can therefore be safely considered as a second-order parameter in the column density calculation compared to other more impactful assumptions in the model. See for example our fiducial no-BW model (solid blue line) using $h = R_d/8$, compared with $h = R_d/15$ (dotted red line). Therefore, the effect of the BW, as seen in Figure 3, is larger than the effect of changing disc thickness.

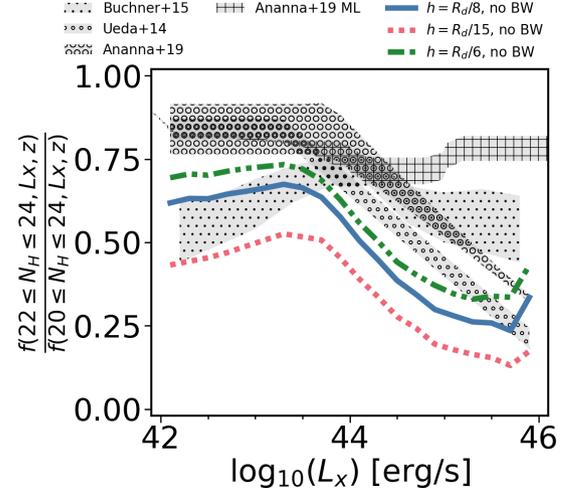

**Figure A1.** Host galaxy CTN obscured sources for the fiducial host galaxy model without BW modifying the prescription of the scale height at redshift $z = 2.4$. We compare the disc heights $h = R_d/6$, $h = R_d/8$ and $h = R_d/15$. Observations as in Figure 3.

## APPENDIX B: DEPENDENCE ON LIGHT CURVE

To determine the dependence of AGN obscured fractions on luminosity, it is necessary in the first place to calculate and modify the AGN luminosity of all sources in the mock catalogue. As mentioned in the main text, we compute the bolometric luminosity from the QSO and Radio accretion rates from GAEA following F20, and then we use Duras et al. (2020) to determine the X-ray luminosity. But throughout the paper, we have been using the peak accretion rate (and therefore the peak bolometric luminosity and peak X-ray luminosity) as the value assigned to each column density. Using another luminosity within the light curve could lead to different results.

The results of this test are shown in Figure B1. In the left panel, CTN fractions decrease when using a random value within the original GAEA light curve, except at higher luminosities. Choosing random X-ray luminosities within the GAEA predicted AGN evolution tends to pick more frequently luminosities lower than the peak, and during the post-peak, more extended phase. This effect is causing many AGN to be selected at luminosities below the $10^{42}$ erg/s limit, thus decreasing the fractions at faint, but not necessarily at high luminosities.

The HQ11-GAEA assumes the Hopkins et al. (2006) light curve. To test the effect of the shape of the light curve on our results, we deploy a slightly different alternative modelling, a *modified* light curve model (*modified* model from now on) characterized by two phases. A first regime is defined by an exponential increase until the galaxy reaches its critical SMBH mass at the peak luminosity,

$$\dot{M}_{\rm BH}(t) = \dot{M}_{\rm BH}^{\rm crit} \exp\left(\frac{t - t_{\rm peak}}{t_{\rm Edd}}\right), \tag{B1}$$

where $\dot{M}_{\rm BH}(t)$ is the accretion rate onto the central SMBH, $\dot{M}_{\rm BH}^{\rm crit}$

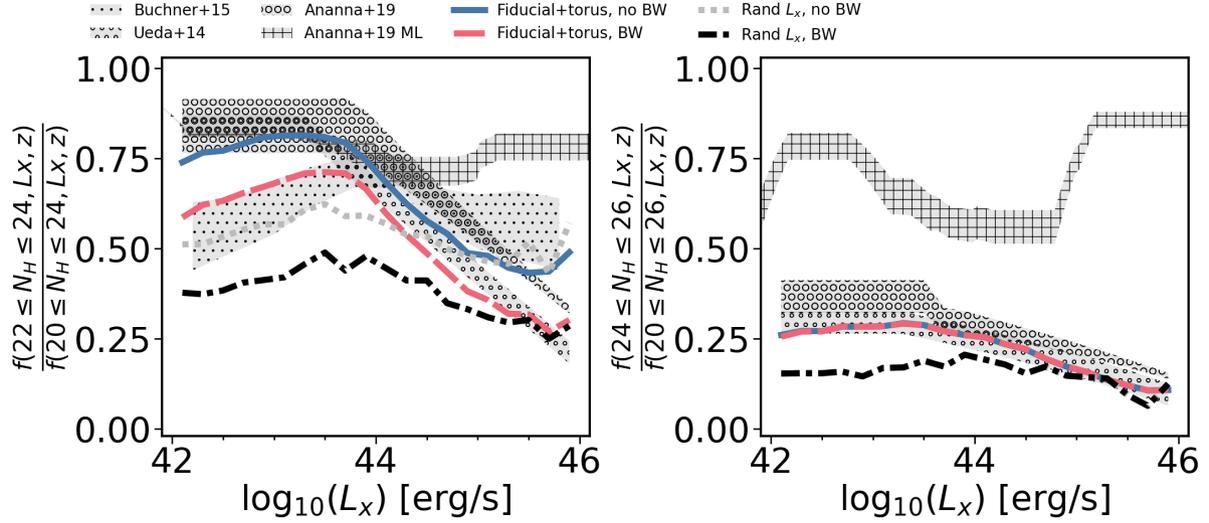

**Figure B1.** CTN (left panel) and CTK (right panel) obscured fractions for the fiducial model without BW (solid blue line), fiducial model with the BW (dashed red line), no BW model but using the randomly picked X-ray luminosity (dotted black line), and the BW model using the randomly picked X-ray luminosity (dash-dotted grey line). Observations as in Figure 3.

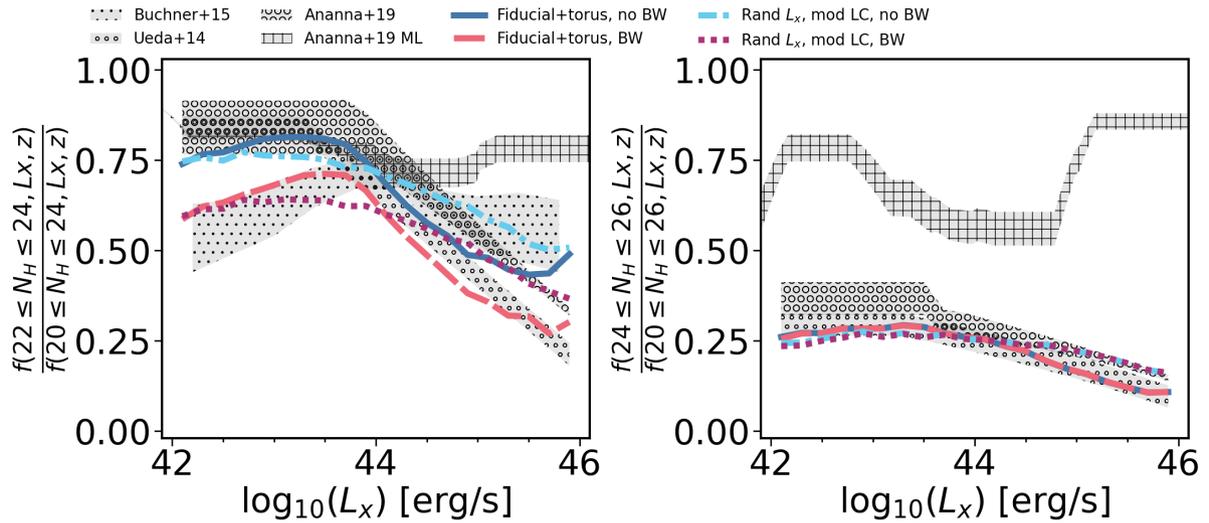

**Figure B2.** AGN obscured fractions depending on the X-ray luminosity randomly selected within the light curve assuming the *modified* light curve, for both with and without the BW at redshift $z = 2.4$. Observations as in Figure 3. **Left panel:** CTN obscured fractions. **Right panel:** CTK obscured fractions.

is the peak accretion rate, $t_{\rm peak}$ is the time corresponding to the peak accretion rate, and $t_{Edd}$ is the Eddington time corresponding to $4.5 \cdot 10^7$ years for our chosen value of the radiative efficiency. Both $\dot{M}_{\rm BH}^{\rm crit}$ and $t_{\rm peak}$ formulas are given by Hopkins et al. (2007). In our test, the $\dot{M}_{\rm BH}^{\rm crit}$ is given by GAEA, but we opt for a random $t_{\rm peak}$ within the possible GAEA timesteps.

A second regime follows a power-law decline as defined by Hopkins et al. (2006),

$$\dot{M}_{\rm BH} = \frac{\dot{M}_{\rm BH}^{\rm crit}}{1 + \left|\frac{t-t_{\rm peak}}{t_{\rm Edd}}\right|^2}, \tag{B2}$$

which is the same equation used in HQ11-GAEA for the post-peak phase (Eq. (14) of F20).

The above model, although still very similar to the original in GAEA, tends to produce more extended curves rather than sharp peaks, with more long-lasting pre- and post-peak phases. We present the results using both the GAEA original and modified models in Figure B2, for models with and without BW. X-ray luminosities are assigned at random within the *modified* model. The overall shapes and normalizations of the predicted CTN fractions are very similar to our GAEA ones, with only a slight decrease of the fractions of obscured AGN in both the CTN and CTK regimes at lower luminosities and an increase at higher luminosities, flattening the fractions and slope. Choosing random X-ray luminosities within the extended AGN curves tends to pick more frequently luminosities lower than the peak, especially during the (longer) post-peak phase. This selection again causes many AGN to fall below the $10^{42}$ erg/s cut, thus decreasing the fraction of faint AGN.

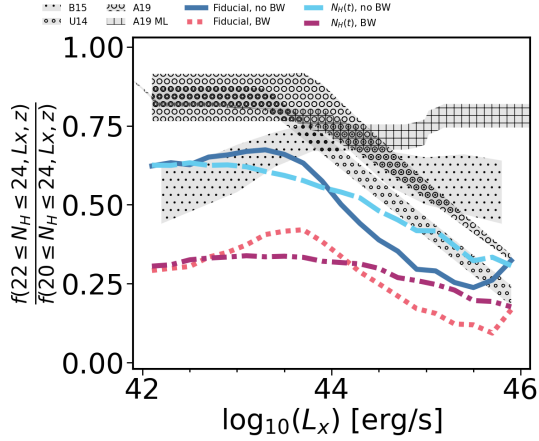

**Figure B3.** Host galaxy CTN obscured fractions for the fiducial model without BW (solid blue line), fiducial model with the BW (dotted red line), no BW model but using the randomly picked X-ray luminosity with a time dependent column density (dashed cyan line), and the BW model by using the randomly picked X-ray luminosity with a time dependent column density (dash-dotted pink line) at redshift $z = 2.4$. Observations as in Figure 3.

### B1 Dependence on time-varying HI column density

In this Appendix we test the impact on the AGN obscured fraction of allowing the gas mass, and thus the $N_H$, to vary within the relatively short timescale of the AGN light curve. In other words, we here test a variant of our reference model in which we include an efficient AGN feedback and/or star-formation rate consumption that can significantly and rapidly decrease the initial gas mass. To this purpose, we follow Granato et al. (2004) who suggest that the gas mass can in some instances decrease exponentially due to AGN feedback, and assume that the $N_H$ column density evolves with time as

$$N_H = N_{H, \text{peak}} \cdot \exp\left(-\frac{t - t_0}{\tau}\right), \tag{B3}$$

where $N_{H,\text{peak}}$ is the column density at the peak luminosity calculated from Equation 4 or 5, $t_0$ is the start of the modified light curve, and $\tau = 2$ Gyr to roughly mimic the time behaviour predicted by Granato et al. (2004). The time $t$ in Eq B3 is the time, within the light curve, corresponding to the $L_X$ selected at random for each source as discussed previously.

We show our results in Figure B3 for our fiducial host model. When comparing with the outputs in Figure 3, it is clear that the predicted fractions with a strongly decreasing gas mass are close to the ones with constant $N_H$, as expected given the relatively short AGN lifetimes. Therefore, for simplicity we continue assuming a constant $N_H$ throughout the lifetime of the AGN as this time-dependent evolution of the column density is not going to heavily impact our results.

### B2 Modified light curve with short-lived torus

Once we have a working model with random X-ray luminosity in our *modified* light curve model, we can test other physically motivated prescriptions for AGN obscuration. As an example, we can test the impact on our results of a potentially short-lived torus component, appearing only during a specific portion of the AGN lifetime. In order to test this idea, we build a toy model where the torus is only present during the pre-peak phase, but rapidly disappears, due to, e.g., AGN feedback and/or gas consumption, during the post-peak phase. In this scenario, if the source is selected in the case of pre-peak, we include the column density coming from our fiducial torus model. On the contrary, if the source is in the post-peak phase, we do not include the torus. We compare the predictions of this toy model against our fiducial model from Figure 7 in Figure B4. It is interesting to see that with this new prescription for a shorter appearance of the torus (dot-dashed, purple and blue, dotted lines in Figure B4), the fraction of obscured AGN decreases, in particular the CTK AGN now reduce to ∼13%, which is noticeably below any of our comparison data sets. This suppression is also evident in CTN obscuration, where it becomes even more marked when including the BW.

We find that if we go back to a standard Wada torus model (Figure B5), which was predicting a larger fraction of CTK sources than our reference model (Figure 3), we are able to recover a sufficiently high fraction of CTK comparable to the number observed (dot-dashed, purple and blue, dotted lines). We conclude that the features of the specific torus model adopted are degenerate with the lifetime of the torus.

## APPENDIX C: DEPENDENCE OF THE AGN OBSCURED FRACTIONS ON THE GAS FRACTIONS IN THE HOST GALAXIES

The column densities are directly proportional to the amount of cold gas mass $M_{\text{cold}}$ in the host galaxy, we expect a variation of $M_{\text{cold}}$ to have an impact on the implied fractions of obscured AGN. In this Appendix, we replace the gas fraction predicted by GAEA (and self-consistently computed in the model as a balance between cooling, star formation and AGN feedback), with the empirical relations derived from the GOODS-S, GOODS-N and the COSMOS fields sample (Santini et al. 2014). This choice allows us to check the impact on the predicted NH distributions when varying the underlying gas fraction in the model. The analytic fit by Santini et al. (2014) suggests an SFR-dependent total gas mass of the form

$$M_{\text{gas}} = \frac{f_{\text{gas}}}{1 - f_{\text{gas}}} M_\star, \tag{C1}$$

with gas fractions calculated as

$$\log f_{\text{gas}} = \alpha + \beta * (\log M_\star - 11). \tag{C2}$$

The variables $\alpha$ and $\beta$ depend on the SFR of the galaxy and can be found in Table 1 of Santini et al. (2014).

Figure C1 compares our reference model with cold gas masses from GAEA with the ones using Santini et al. (2014). The GAEA models with and without BW (solid blue and red dot-dashed lines, respectively) have broadly similar predictions for the fractions of CTN AGN to the models assuming the cold gas masses from Santini et al. (2014) with and without BW (dashed green and dash-dotted yellow lines, respectively). Despite relatively minor differences, the mean CTN fractions are similar, proving that the gas fractions from GAEA are sufficiently reliable and not biasing our core results.

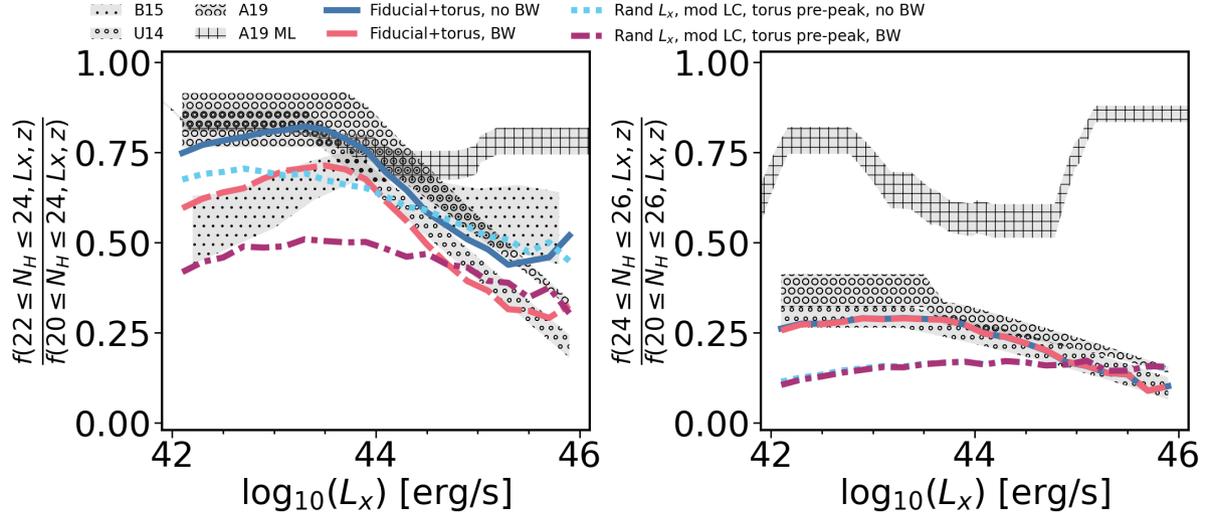

**Figure B4.** AGN obscured fractions depending on the X-ray luminosity of the fiducial galaxy model compared against the random X-ray luminosity within the *modified* light curve, for both BW and no-BW, assuming that the fiducial torus model only appears in the pre-peak phase at redshift $z = 2.4$. Observations as in Figure 3. **Left panel:** CTN obscured fractions. **Right panel:** CTK obscured fractions.

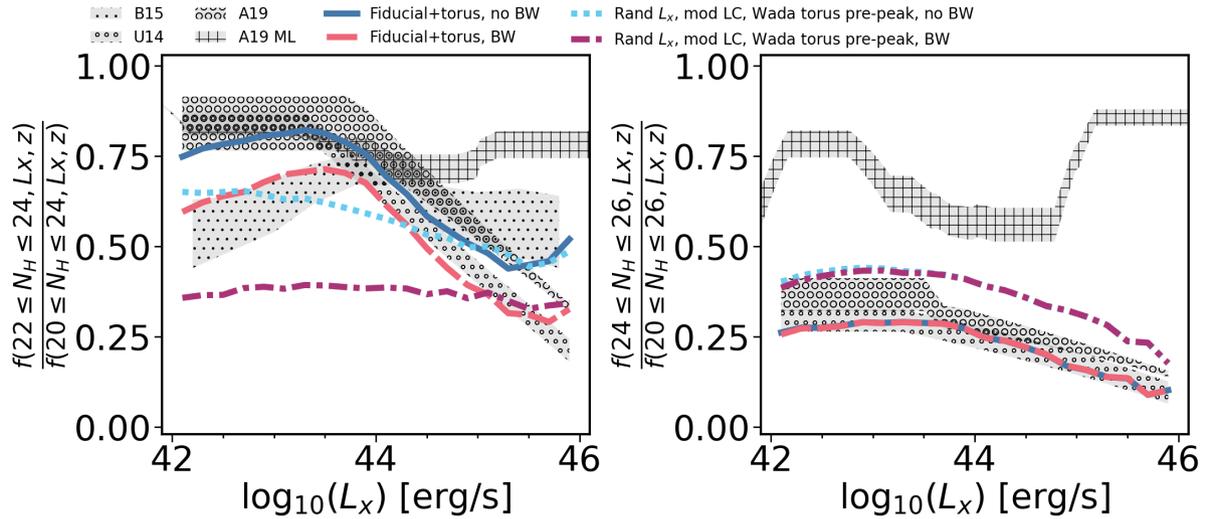

**Figure B5.** AGN obscured fractions depending on the X-ray luminosity of the fiducial galaxy model compared against the random X-ray luminosity within the *modified* light curve, for both BW and no-BW model, assuming that the Wada torus model only appears in the pre-peak phase at redshift $z = 2.4$. Observations as in Figure 3. **Left panel:** CTN obscured fractions. **Right panel:** CTK obscured fractions.

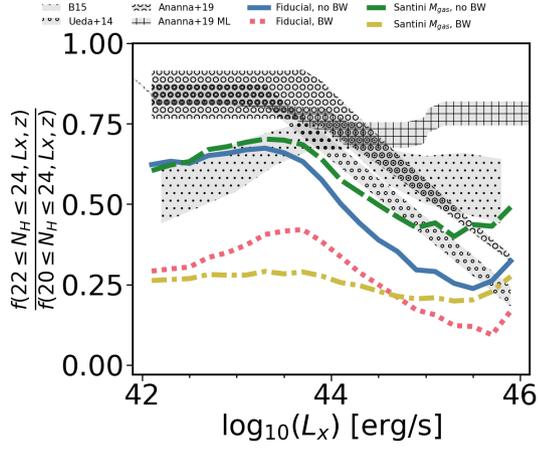

**Figure C1.** CTN obscured fractions of the host galaxy using the empirical relation from Santini et al. (2014) to estimate cold gas fractions at redshift $z = 2.4$. We compare with the original GAEA cold gas fractions as in our fiducial host model. Observations as in Figure 3.